\documentclass[]{aa}

\usepackage{xcolor}

\usepackage{amsmath, amssymb}
\usepackage{booktabs, makecell}
\usepackage{color}
\usepackage{graphicx}
\usepackage{hyperref} 
\usepackage{lscape}
\usepackage{microtype}
\usepackage{mwe}
\usepackage{natbib}
\usepackage{rotating} 
\usepackage{subfig}
\usepackage{threeparttable}
\usepackage[varg]{txfonts}
\usepackage{url}
\usepackage{xspace}
\usepackage{float}
\usepackage{longtable}

\def\cxo{{Chandra}\xspace}
\def\ero{{eROSITA}\xspace}

\def\gai{{Gaia}\xspace}

\def\heo{He{\sc I}\xspace}
\def\het{He{\sc II}\xspace}

\def\srgero{{SRG/eROSITA}\xspace}

\def\tes{{TESS}\xspace}

\def\xmmn{{XMM-Newton}\xspace}

\newcommand\fergs{\ensuremath{\mathrm{erg}\,\mathrm{cm}^{-2}\,\mathrm{s}^{-1}}\xspace}

\newcommand\lx{\ensuremath{\mathrm{erg}\,\mathrm{s}^{-1}}\xspace}

\newcommand\porb{\ensuremath{P_{\rm orb}}\xspace}

\begin{document}

\title{Cataclysmic variables from SRG/eROSITA: new systems from eRASS1}

\author{A.D. Schwope\inst{1}
\and
J. Brink\inst{1,2}
\and
D.A.H. Buckley\inst{3,4,5}
\and
V.~A.~C\'uneo \inst{1}
\and
S. Friedrich\inst{6}
\and
S. Hernández-Díaz\inst{7}
\and
K. Knauff\inst{1}
\and
J. Kurpas\inst{1}
\and
G. Lamer\inst{1}
\and
S.B. Potter\inst{3,8}
\and
K.G. Pradeep\inst{1,2}
\and
M.R. Schreiber\inst{9}
\and
B. Stelzer\inst{7}
\and
J. R. Thorstensen\inst{10}
\and
D. Tubin-Arenas\inst{1}
}
\institute{Leibniz-Institut für Astrophysik Potsdam (AIP), An der Sternwarte 16, 14482 Potsdam, Germany \email{aschwope@aip.de}
\and
Institute for Physics and Astronomy, University of Potsdam, Karl-Liebknecht-Str. 24/25, 14476 Potsdam, Germany
\and
            South African Astronomical Observatory, PO Box 9, Observatory Road, Observatory 7935, Cape Town, South Africa
\and
            Department of Astronomy, University of Cape Town, Private Bag X3, Rondebosch 7701, South Africa
\and
            Department of Physics, University of the Free State, PO Box 339, Bloemfontein 9300, South Africa
\and
 Max Planck Institute for Extraterrestrial Physics, Gie{\ss}enbachstra{\ss}e 1, 85748 Garching, Germany
\and  
Institut für Astronomie und Astrophysik, Eberhard Karls Universität Tübingen, Sand 1, 72076 Tübingen, Germany
\and
            Department of Physics, University of Johannesburg, PO Box 524, Auckland Park 2006, South Africa
\and
Departamento de F\'isica, Universidad T\'ecnica Federico Santa Mar\'ia, Av. España 1680, Valpara\'iso, Chile
 \and
            Department of Physics and Astronomy, Dartmouth College, Hanover NH 03755, USA
}
\date{Received ; accepted}

\abstract{The X-ray all-sky surveys performed with \ero on the Spektrum-Roentgen-Gamma (SRG) mission offer a unique opportunity to generate large samples of all kinds of X-ray emitters. Here we focus on a minority population among the X-ray point sources, the cataclysmic variables (CVs).}
{We aim to generate large samples of CVs selected from the \srgero surveys to address fundamental questions about binary evolution, the role of magnetic fields in CV evolution and their contribution to the Galactic Ridge X-ray Emission (GRXE).}
{We have trained a Random Forest (RF)  classifier based on combined X-ray and optical properties to select CV candidates from the first \ero X-ray all-sky survey (eRASS1). Follow-up spectroscopy to securly identify the objects was performed on a subset of the selected candidates using telescopes at the Northern and the Southern hemisphere. Newly identified CVs were further analyzed to determine their likely subtype with the help of dedicated follow-up photometry, spectroscopy and archival resources.}
{We have identified 156 CVs of almost all subtypes covering a wide range of distances (between 170\,pc and several kpc), absolute magnitudes ($G=4.5 - 12$) and X-ray luminosities ($\log L_X (0.2-2.3 \mbox{keV})=29.6 - 33.3$\,\lx). For most objects, the nature as a CV is reported here for the first time. For 40 objects periods were determined, which were regarded as their likely orbital periods. These range from 79.2 min, at the CV minimum period,  to almost 22 hours. Seven objects were found to be eclipsing, a lower limit to the actual number due to the current lack of dedicated follow-up observations. A further five objects show cyclotron or Zeeman features, that allowed to determine their field strengths. A soft component was found in 14 CVs, all were regarded being magnetic. The sample comprises X-ray luminous objects, excellent candidates for being new Intermediate Polars, that are thought to be main contributors to the GRXE. The new method of selecting CVs from eRASS1 was found to be superior to other methods described in the literature and the new CVs were used to further refine the method.}
{This work demonstrates the feasibility to construct comprehensive CV samples through follow-up spectroscopy of \srgero X-ray sources. It has also demonstrated the potential to uncover rare objects which are of scientific interest on their own.}


\maketitle
\nolinenumbers
\section{Introduction}\label{s:intro}
Cataclysmic variables (CVs) are compact binaries with an accreting white dwarf (WD) and a Roche-lobe filling donor or secondary star \citep{warner95}. Most CVs host main-sequence donors, but symbiotic stars  (SySts) with evolved donors and ultracompact binaries (UCBs, AM CVn stars) with a compact degenerate secondary might also be included in the CV family. CVs are the most numerous accreting compact binaries in our own galaxy. They represent important laboratories for studying compact binary evolution \citep{belloni_schreiber23} and are important accretion laboratories, be it in disks or, for the magnetic CVs, in accretion columns. They serve as excellent laboratories for the planned space-based gravity wave observatory LISA and subcategories are thought to host the progenitors of SN\,Ia. The summed X-ray signal of all CVs is assumed to be a main contributor to the famous Galactic Ridge X-ray emission \citep[GRXE, ][]{worrall+82, suleimanov+22}.

\begin{table*}[t]
\begin{center}
\caption{Facilities used for spectroscopic identification of \ero-selected CV candidates from eRASS1\label{t:facil}}
\begin{tabular}{llccccr}
\hline \hline
Telescope & Instr. & Grism & Coverage & Dispersion & Resolution & \#objects \\
 & & & \AA & \AA/pixel & \AA & \\
\hline
ESO/NTT & EFOSC2 & \#6    & 3860--8070 & 2.1 & 16  & 138\\
MDM 2.4m & OSMOS & Blue   & 3975-6780  & 0.7 & 3.3 &  14\\
NOT     & ALFOSC & \#7    & 3680--7140 & 1.7 &  7  &   2\\
SAAO 1m Lesedi & Mookodi &        & 4000--8000 &  3.9  &  6   &  17\\
SALT    & RSS    & PG0700 & 3200--6760 & 1.2 & 7.8 &   4\\
\hline
\end{tabular}
\end{center}
\end{table*}

Despite their general importance, the population properties (space density, scale height, X-ray luminosity function) and evolutionary pathways of both magnetic CVs (MCVs) and non-magnetic CVs (nMCVs) remain poorly constrained. The markedly different orbital-period distributions of MCVs, which do not show an orbital period gap, and nMCVs, which show a pronounced orbital period gap between 147 min and 191 min, respectively, suggest that the two populations follow distinct evolutionary pathways \citep{schreiber+24}. One possible explanation is the evolutionary scenario proposed by \citet{schreiber+21}, in which strong magnetic fields emerge only late in the evolution of the white dwarf, fundamentally altering the subsequent evolution of the binary. While this framework successfully explains several observed properties of magnetic CVs, a quantitative population model 
of both magnetic and non-magnetic systems is still lacking.

Progress was hampered in the past because of small and/or biased and incomplete samples. CVs were either found as optically variable objects, or blue excess objects, as counterparts of X-ray sources or just serendipitously mimicking an AGN in the early incarnations of the SDSS project \citep[see e.g.~the discussion in ][]{gaensicke05}. 

Fortunately, the situation is changing now through the advent of sensitive all-sky surveys in the optical and the X-ray regime. All CVs are X-ray emitters with measured X-ray luminosities between approximately $10^{29}$\,\lx and  $10^{34}$\,\lx \citep[$0.2-2.3$ keV,][]{schwope+24b}. The \srgero all-sky surveys \citep[eRASS,][]{predehl+21, sunyaev+21} thus allow to conduct X-ray surveys in a volume with a radius of a few hundred parsec to a few kpc to sample the relevant subclasses, their sky positions, fluxes, and luminosities. \cite{schwope+24b} have shown, that volume-limited samples of most types of CVs can be generated with a radius of 500 pc and completeness fraction of 90\% or higher. Only the period bouncers and other low mass-transfer-rate objects such as the LARPs \citep[Low Accretion Rate Polars, ][]{schwope+02b} have smaller survey volumes of 200 to 300 pc. To put those numbers into perspective: the only existing most complete volume-limited CV sample has a radius of 150\,pc and contains 42 objects at an estimated completeness of 80\% \citep[][]{pala+20}. The 300\,pc 'Golden sample' of \cite{inight+21} comprises 151 objects and has an estimated completeness of about 50\%.

In addition to the X-ray route of finding new CVs, photometric surveys like the ZTF and spectroscopic surveys like DESI and SDSS open up complementary paths to establish the desired complete samples \citep[see e.g.][]{szkody+21, hou+26}. The SDSS has meanwhile obtained several hundred spectra of CVs \citep{inight+23a, inight+23b, inight+25} that were obtained for various reasons (as likely QSOs or as strongly variable objects). Meanwhile, systematic spectroscopic identification campaigns are underway within the SDSS to establish the CV population \citep{kollmeier+26, schwope+24, brink+26, hernandez-diaz+26}. Furthermore, while finishing this paper, the 4MOST facility will start its 5 years survey to perform the deepest and most comprehensive X-ray follow-up survey of \srgero candidates, which includes a dedicated search for CVs \citep{dejong+19, chiappini+19}.   

The current paper is the first to describe a larger sample of new objects from the \srgero surveys. Previous publications based on \ero surveys have reported small samples in selected areas or individual objects that attracted attention due to their exceptionally high or low brightness, their pronounced X-ray variability, or their proximity \citep{avakyan+25,balman+24,bobakov+25,munoz-giraldo+25,kolbin+24, rodriguez+25,schwope+22a, schwope+22b, schwope+24}. Here we describe the first results of comprehensive spectroscopic follow-up observations of CV candidates that were found during the first \ero survey. Full X-ray information for all objects published here can be retrieved from the DR1 catalog of \ero X-ray sources published by \cite{merloni+24}. 

The paper is organized in the following way. In Section \ref{s:ts} the target selection and in Section \ref{s:obs} the observations are described. Section~\ref{s:ana} informs about the analysis of our new spectra and of available photometry, both from dedicated observations and from archival data. Section \ref{s:res} informs about the main results from the spectroscopic observation campaigns. The main part of the paper is finished with a discussion and conclusion in Section ~\ref{s:disc}. In the Appendix we firstly give an overview over the graphical products that were inspected to confirm a CV identification and determine its subtype (Sect.~\ref{a:cvprop}). We give a description of the full list of CVs with all relevant parameters in Section ~\ref{a:cvcat}. The full list is published online only and is available at CDS, but an abridged version is also listed in Section ~\ref{a:cvcat}. During our spectroscopic campaigns not only CVs but also a few other X-ray sources were identified, mostly AGN. These are also reported in the Appendix (Section ~\ref{a:other}). In the following we refer to individual objects abbreviating their full IAU names to Jhhmm$\pm$dd.

\section{Target selection for spectroscopic identification}\label{s:ts}
CVs are a minority population in the X-ray sky. Only about every 1000th source in the eRASS catalogs might be a CV. The selection of CVs thus is not straightforward and candidate CV samples are prone to contamination by stars or AGN. Building a reliable CV sample from an X-ray catalog therefore has to rely on spectroscopic identification and confirmation.

All targets that were chosen for spectroscopic observations were selected from a preliminary version of the \ero catalog of the first X-ray all-sky survey (eRASS1, pipeline version c946), but all sources considered here also appeared in the official DR1 release of the first eRASS. Given the data sharing agreement between the Russian and the German \ero collaborations \citep{merloni+24}, most of our targets have southern declinations. We considered X-ray point sources only and applied a flux limit of $1 \times 10^{-13}$\,\fergs in the $0.5-2$\,keV band for this exploratory study in order to work with very secure X-ray detections only. This left 107396 sources for further analysis. The flux limit is significantly deeper than in the earlier ROSAT study which was limited to $2.4\times 10^{-12}$\,\fergs \citep{schwope+02}. CVs populate very special regions of color-magnitude and color-color-diagrams involving X-ray and optical colors \citep[see the diagnostic diagrams in][and in Sect~\ref{s:landsc} below]{schwope+24b}, which allows to efficiently search for new CV candidates. We compiled a list of currently known CVs from various input sources, e.g.~from \cite{ritter_kolb03}, the Washington open CV catalog\footnote{\url{https://depts.washington.edu/catvar/index.html}} and similar collections, with measured X-ray flux from either \ero, \xmmn, or \cxo. Using this list of 642 CVs as input we developed a machine learning classifier based on a combination of Gaia EDR3\footnote{https://www.cosmos.esa.int/web/gaia/dr3} and X-ray data. Parameters used in training and classification were the X-ray flux, $BP-RP$ color, $G$-band magnitude, proper motion,  distance \cite[$r_{\rm geo}$ from ][was used]{bailer-jones+21}, and \gai variability. The trained classifier learns to map the trends in these parameter spaces into a CV likelihood ($P_{\rm CV}$). The $P_{\rm CV}$ in turn forms the 'prior-probability' for the Bayesian crossmatch tool NWAY\footnote{\url{https://github.com/JohannesBuchner/nway/}}, which is used to find the most reliable \gai counterpart of the eROSITA source and further rank them based on their likelihood to be a CV. At this stage also the X-ray to positional offset is taken into account \citep{salvato+18}. This process revealed several hundred potential counterpart objects for the X-ray sources. Among those were known CVs, that were discarded from further consideration. All the chosen candidates had Gaia distances and most had secure optical counterparts. Eventually about 200 candidate objects were proposed for spectroscopic follow-up observations at northern and southern facilities to confirm or reject their tentative identification as a CV. 
\begin{figure}[t]
\resizebox{\hsize}{!}{\includegraphics[clip=]{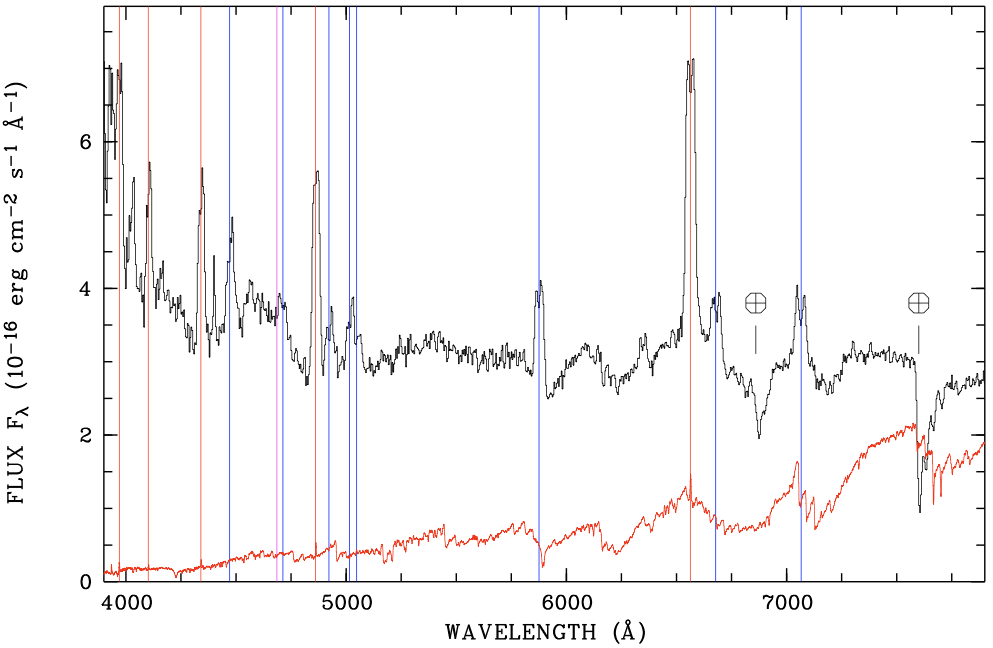}}
\caption{Discovery spectrum of J0356$-$44, a dwarf nova (DN) at 170pc distance. The spectrum was obtained with EFOSC2 at the NTT. Vertical lines indicate main emission lines (H-Balmer lines with red, \heo\ with blue, and \het\ with magenta color). The atmospheric A- and B- absorption bands at 6860\,\AA\ and 7600\,\AA\ are also indicated. Shown with red color is a scaled M4 dwarf star template spectrum from \cite{kesseli+17}. \label{f:j0356_ntt}
}
\end{figure}

\begin{figure}[t]
\resizebox{\hsize}{!}{\includegraphics{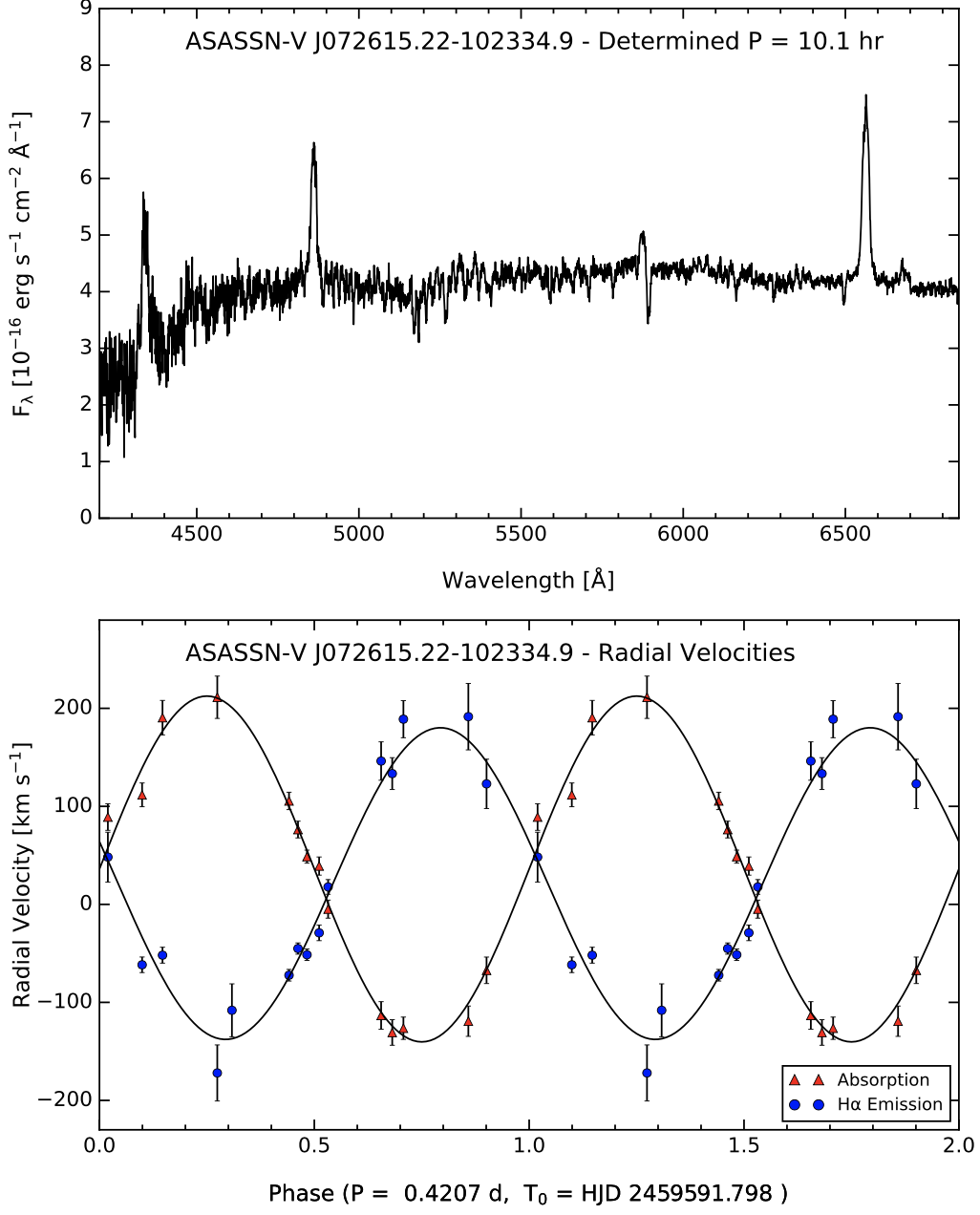}}
\hfill
\caption{Identification spectrum obtained with the MDM 2.4m (top panel) and radial velocity results of phase-resolved spectroscopy (bottom panel) of the new NL J0726$-$10.}
\label{f:0726specrv}
\end{figure}

\section{Observations}\label{s:obs}
\subsection{Optical low-resolution spectroscopy}
 Spectroscopic follow-up for identification of \ero-selected CV candidates was performed with 1$-$4 m class telescopes. Most spectra were obtained with the ESO/NTT during three runs in October 2022, February/March 2023 and July 2023 (10 nights in total allocated to programs 110.24DX and 111.24Q0). 
Additional low-resolution spectra were obtained with the MDM Observatory 2.4m telescope at Kitt Peak, the Nordic Optical Telescope (NOT) at La Palma, with the SALT, and with other SAAO telescopes. At SAAO, spectroscopic observations were primarily obtained using the Mookodi instrument mounted on the South African Astronomical Observatory (SAAO) 1.0 m Lesedi telescope \citep{erasmus+24}. The wavelength calibration was achieved using arc line spectra (e.g.~HeAr at the ESO site, Xenon Pen-Ray arc lamps at SAAO). Standard star spectra chosen from the ESO list\footnote{\url{https://www.eso.org/sci/observing/tools/standards/spectra.html}} were observed regularly for flux calibration. An overview of the facilities used, the main spectral characteristics, and the number of objects that were observed from each site is given in Table \ref{t:facil}. 

For 11 of our CV candidates time-resolved spectroscopic follow-up observations with the MDM 2.4m telescope could be performed by one of us (JT). These data were thus not only used for identification; for some of the newly identified objects these data revealed the binary orbital period of the systems.

\begin{figure*}[t]
\resizebox{0.5\hsize}{!}{\includegraphics[clip=]{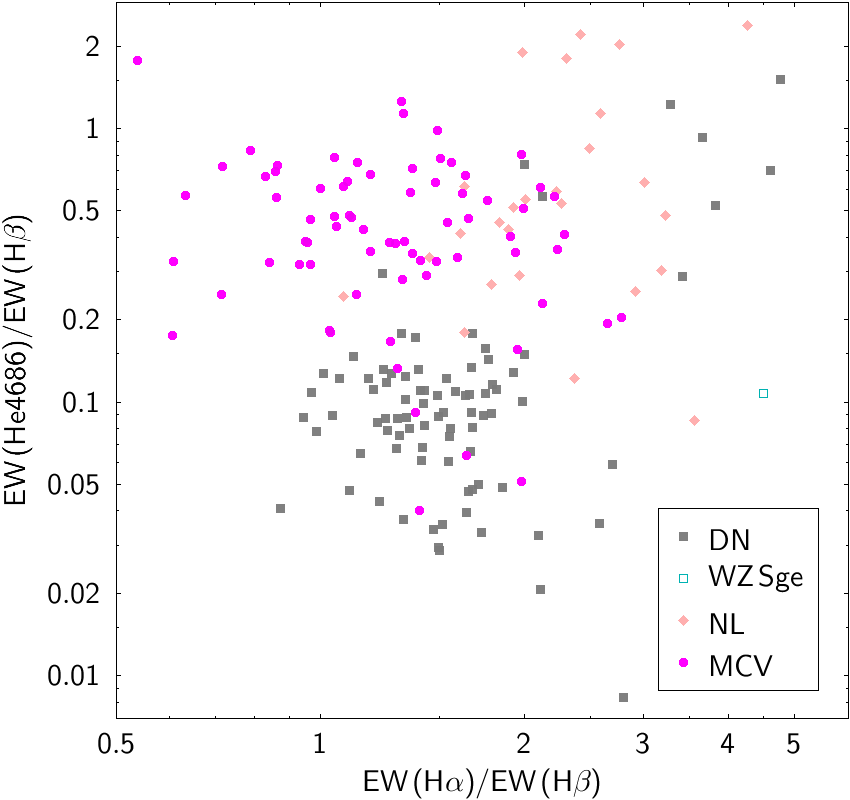}}
\resizebox{0.5\hsize}{!}{\includegraphics[clip=]{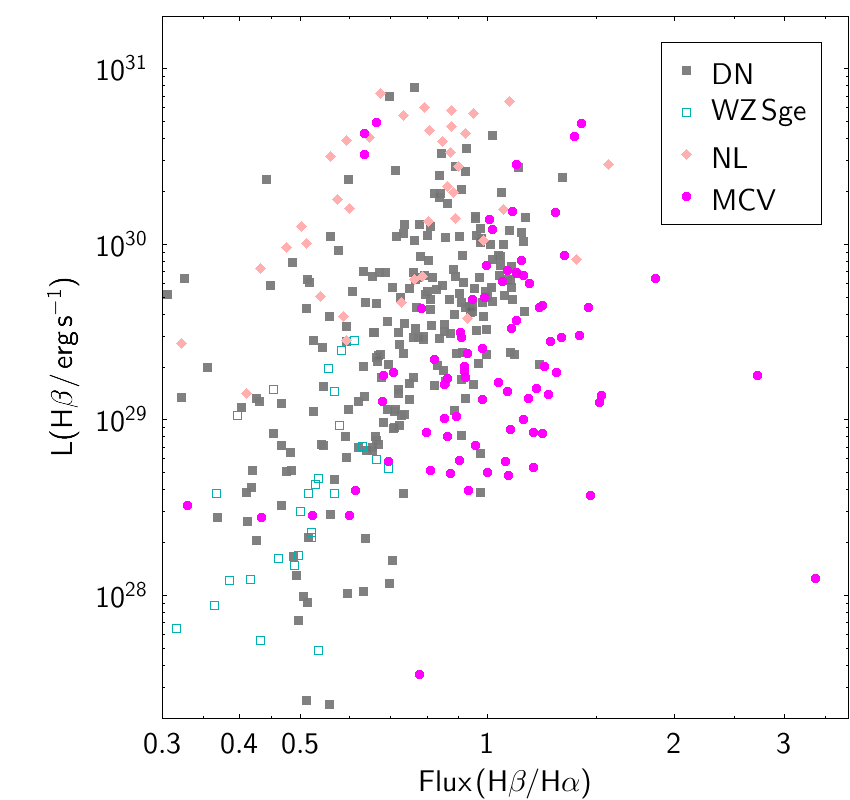}}
\caption{Line diagnostics in CVs presented by \cite{inight+25}. The left panel shows equivalent width ratios of \het, Balmer H$\beta$ and H$\alpha$, the right panel shows the H$\beta$ luminosity as a function of the flux ratio of the two main Balmer lines. 
\label{f:lines}
}
\end{figure*}

\subsection{Time-resolved photometry}
For new CVs identified through spectroscopy, attempts were made to obtain time-resolved photometry to determine periodic brightness variations with the main aim to estimate or measure the binary orbital period. Mostly, observations were organized from three sites, the SAAO (Sutherland site), with the SPECULOOS network of 1m-telescopes in Chile \citep{delrez+18}, and with \tes \citep{ricker+15}. At SAAO, photometric data were acquired using the Mookodi instrument in imaging mode, as well as with the SHOC instrument \citep{coppejans+13} on the SAAO 1.0 m telescope. Both instruments provide a range of filters; however, most observations were conducted without filters to maximize throughput and achieve higher temporal cadence. The SPECULOOS telescopes at the Paranal observatory were used with $g$ or $r$ filters.

SPECULOOS and SAAO data were reduced with standard procedures (dark and bias subtraction, and flatfielding) and differential photometry was performed for our target stars. SAAO photometry was obtained for 24 objects, with SPECULOOS we observed 27 objects, the target lists of SAAO and SPECULOOS were partially overlapping so that in total 47 objects were observed photometrically. 

The \tes archive is constantly growing and new CVs from this program were proposed for high cadence observations in cycles 8 and 9. For the current paper we constrained ourselves to include only data up to sector 96 (approximately end of 2025). Until then, \tes observed 24 of the objects presented here. The \texttt{lightkurve} \citep{2018ascl.soft12013L} Python package was used to extract light curve data from both official \tes and community developed pipelines of our targets, followed by a Lomb-Scargle periodogram analysis  {(\citealt{Lomb}, \citealt{Scargle}), a widely used technique for identifying periodic signals in unevenly sampled observations. Phase-folded light curves were then generated for significant peaks and further inspected. For finally accepted periods a bootstrapping analysis was performed with 3000 iterations from which the mean, the median and the standard deviation of the period was determined, the latter was used as the error of the period (see e.g. \citealt{hernandez+25}). 

\subsection{Long-term optical light curves from archival photometry}
Long-term light curves were constructed by combining archival photometric data from various time-domain surveys. We used data from the Catalina Real-time Transient Survey \citep[CRTS;][]{drake+09}, the Asteroid Terrestrial-impact Last Alert System \citep[ATLAS][]{tonry+18}, the All-Sky Automated Survey for Supernovae (ASAS-SN; Shappee et al. 2014, Christy et al. 2023), and the Zwicky Transient Facility \citep[ZTF,][]{masci+19}. Together, these surveys provide optical coverage spanning more than a decade for many of our candidates. This extensive coverage allows us to investigate the long-term variability, identify outburst events or transitions from high to low accretion states and vice versa. Such long-term light curves were used to determine the subtype and subclass of the new CVs. They are displayed for each object in the overview graphs that were constructed for each object (see Fig.~\ref{f:1diag} and Sect.~\ref{s:data}).

\section{Analysis}\label{s:ana}
The basis of our compilation of CVs is a low-resolution spectrum as shown from our identification program in Figs.~\ref{f:j0356_ntt} and \ref{f:0726specrv}. The spectral analysis was augmented by locating the objects with their measured or inferred parameters in diagnostic diagrams, and by the analysis of time-resolved photometry to determine -- or at least to constrain -- the likely subtype of a CV. We use rather broad sub-classes. Among the non-magnetic CVs we discern between the nova-likes (NL), and the dwarf nova (DN), the latter with the further subclass of WZ Sge systems (WZ). These are of particular interest, because they might be period-bouncers (PBs). Among the MCVs we tried to discern between the intermediate polars (IPs) and the polars. The spectra, the diagnostic diagrams, the finding chart, the used line ratio diagrams, some derived parameters and our likely subclassification are shown object by object in profile sheets which are made available online, one example is shown in the appendix (see Figs.~\ref{f:1diag}).

\begin{figure}[t]
\resizebox{\hsize}{!}{\includegraphics[clip=]{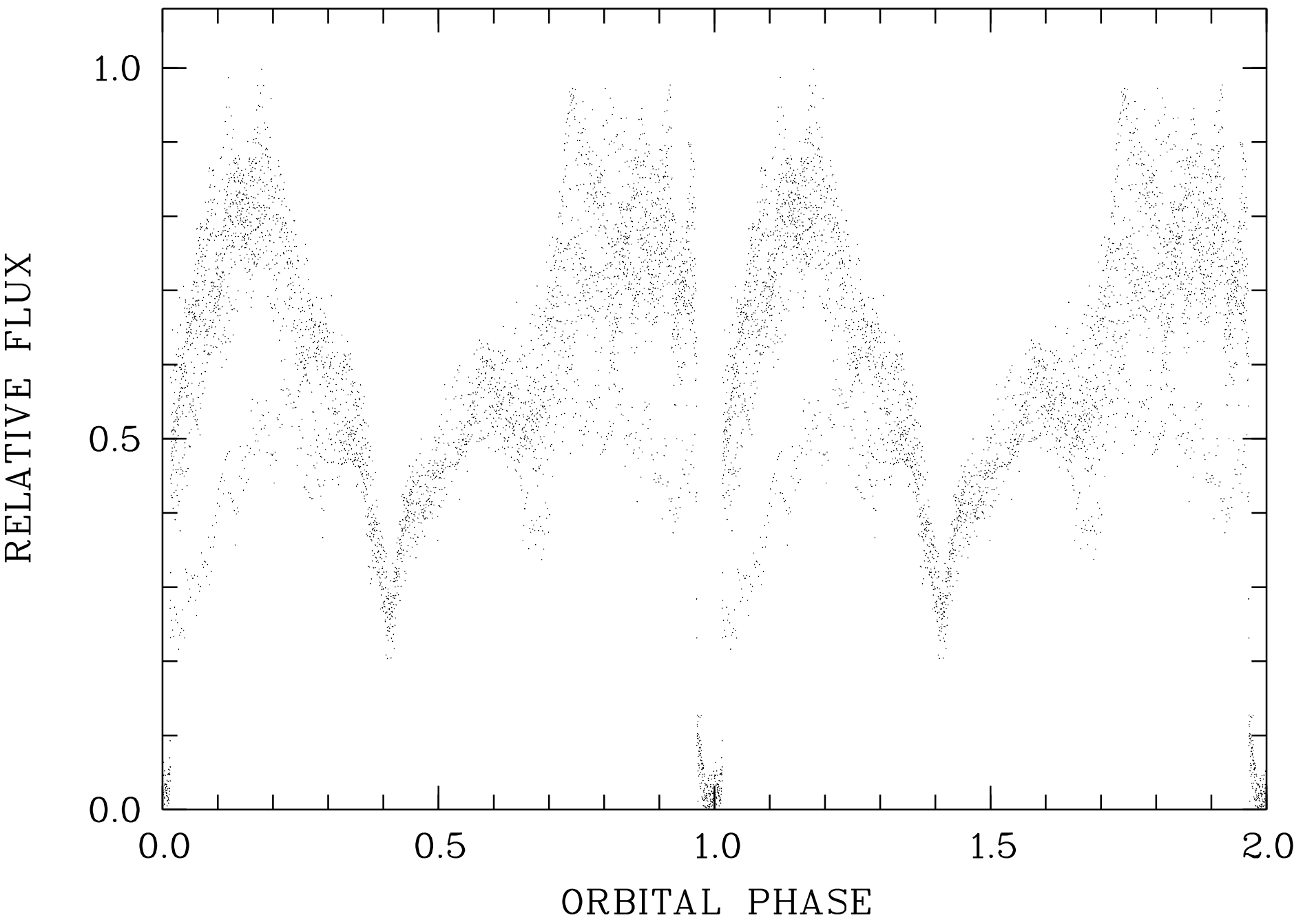}}
\caption{SAAO SHOC phase-folded light curve of J01333$-$81. The original data were folded over the binary orbital period of 79.226 min.
\label{f:lc_j01333}
}
\end{figure}

\subsection{Spectroscopic analysis}
We obtained 192 spectra from 162 unique CV candidates that were subsequently analyzed. For some objects spectra were obtained from two sites. The spectra were visually screened to decide if the object was a CV and, if this was not confirmed, if nevertheless the object was the likely counterpart of the X-ray source. This revealed 156 CVs and five AGN and two possible active stars or CV candidates. The analysis of the confirmed CVs is presented in the main body of the paper; the further X-ray identifications are presented in the appendix (see Sect.~\ref{a:other}).

The identification of a CV as such is based on typical spectral features, which were recently discussed per CV subclass by \cite{inight+23a} and therefore will not be repeated here. It rests on the composition of the emission line spectrum and the shape of the continuum. 

\subsubsection{Line ratio analysis}
All CVs presented here show broad hydrogen Balmer lines, and often also lines of neutral or ionized helium that may originate from accretion disks, hot spots on the disks, accretion streams, accretion curtains, or the irradiated hemispheres of the donor stars. Depending on the binary orbital phase, line profiles and intensities may display strong variability; some line components may even disappear. During our follow-up program we typically obtained just one spectrum at an arbitrary orbital phase; hence the predictive power of our spectra is limited in this regard, but nevertheless informative. We measured, whenever possible, the line parameters of H$\alpha$, H$\beta$, \heo 4471 and of \het 4686. For H$\alpha$ the full width at half maximum (FWHM), the line flux and the equivalent width (EW) were determined per spectrum with a graphical cursor using ESO/MIDAS. For the other lines only the line flux and EW were determined. For all lines with measured flux the emission line luminosity was inferred using the distances $r_{\rm geo}$ derived by \cite{bailer-jones+21}. 

The same spectral line analysis was performed for all CV spectra presented by \citet{inight+25} whose subtype could be determined uniquely by them. We take their classification at face value; the sample with unique classifications comprises 72 MCVs (92 spectra), 30 NL(41 spectra), 175 DN (237 spectra), and 24 WZ Sge-type DN (36 spectra). We use this sample as background information, because all the spectra were obtained with the SDSS, hence have the same data format and the same spectral resolution, which facilitates the analysis. The full analysis of the emission line measurements of uniquely classified CVs from the SDSS will be presented elsewhere, but part of it was used here and two of the diagnostics are therefore shown in Fig.~\ref{f:lines}, the EW ratio $\het/$H$\beta$ over that of H$\alpha$/H$\beta$ and the line luminosity of H$\beta$ vs.~the line flux ratio H$\beta$/H$\alpha$. More of these diagrams that were used by us are shown in Fig.~\ref{f:1diag}.

The data shown in Fig.~\ref{f:lines} show, that an object with EW(\het)/EW(H$\beta$)$>0.2$ is likely an MCV (if EW(H$\alpha$/EW(H$\beta$)$\leq$2), and is likely a NL for  EW(H$\alpha$/EW(H$\beta$)$\geq$2. Objects with EW(\het)/EW(H$\beta$)$<0.2$ are most likely DN. The WZ objects do not occur in the left panel, they typically don't show helium emission lines. The right panel is therefore based on hydrogen Balmer lines only. The H$\beta$ luminosity of NL is highest, that of DN and MCVs typically one order of magnitude lower, but with a large dispersion. The WZ objects are found in the lower left corner of the diagram, i.e.~at $L(\mbox{H}\beta) < 10^{29}$\,\lx and a flux ratio H$\beta$/H$\alpha < 0.6$. 

\begin{figure}[t]
\resizebox{\hsize}{!}{\includegraphics[clip=]{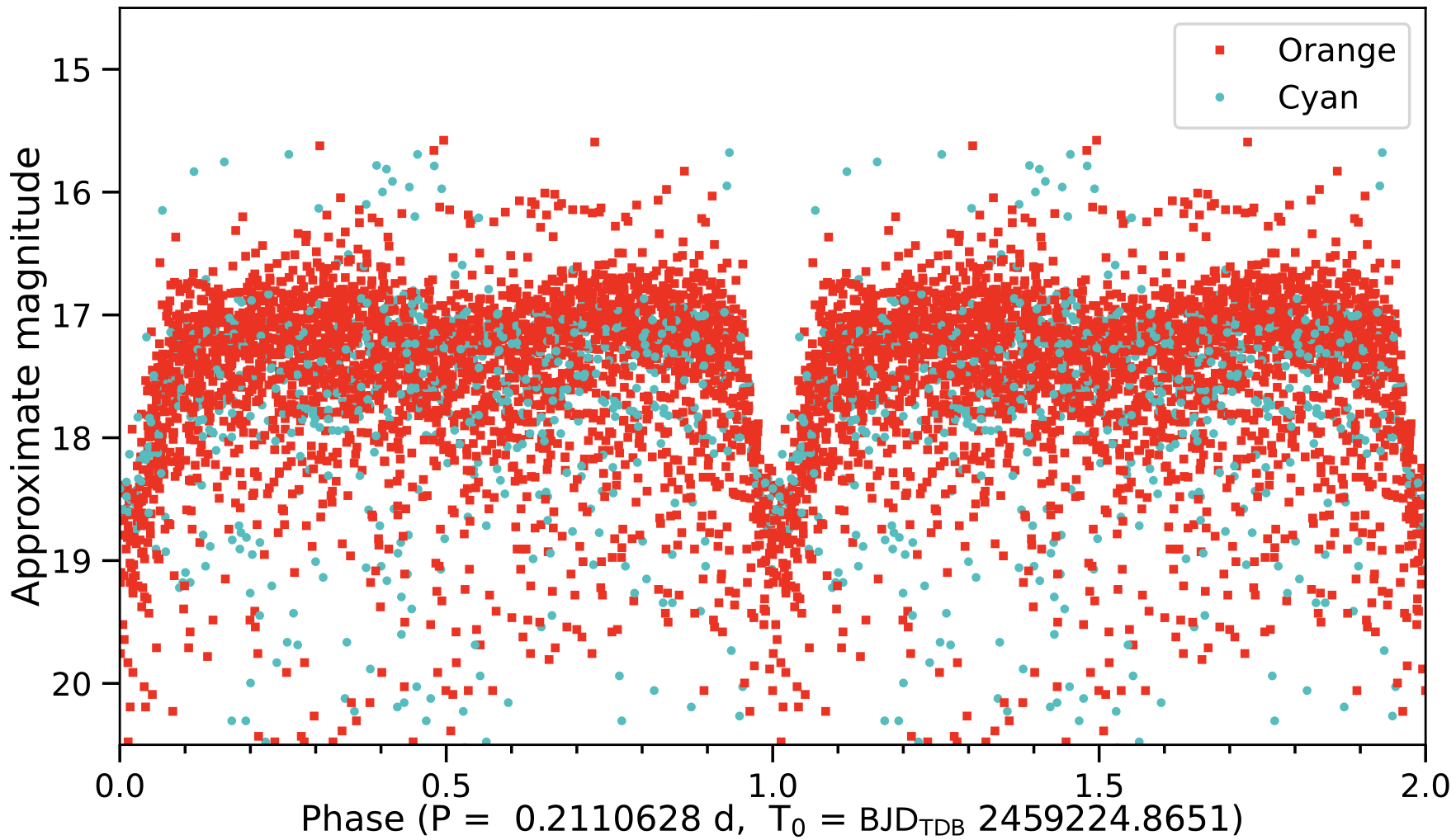}}
\caption{Phase-folded ATLAS light curve of J0715$+$08
\label{f:asn0715}
}
\end{figure}

\subsubsection{Continuum spectrum}
The continuum of a CV might be composed of stellar and non-stellar contributions. For objects accreting at a low level, the stellar photospheres might be recognized, but often the continuum is just rising towards short wavelengths. A distinction between a WD-dominated and a disk-dominated spectrum, which have similar overall shapes and may show broad absorption lines, is possible by locating the object in a \gai-based color magnitude diagram (see Sect.~\ref{s:catval}). WD-dominated objects are located close to the WD sequence and are low mass-transfer systems like WZ-type DN, or polars in their low states, while disk-dominated high-mass transfer objects are located close to the main sequence (see examples in the suite of diagnostic diagrams including the CMD that are published online and for one example Fig.~\ref{f:1diag}). Contributions to the continuum may also come from hot spots on disks or on the WD, or from an accretion curtain, and from the photosphere of the donor. In magnetic objects, cyclotron radiation might shape the continuum and we clearly find two such examples in our sample (J1241$-$60, J1754$-$44).

\begin{figure*}[t]
\resizebox{\hsize}{!}{\includegraphics[clip=]{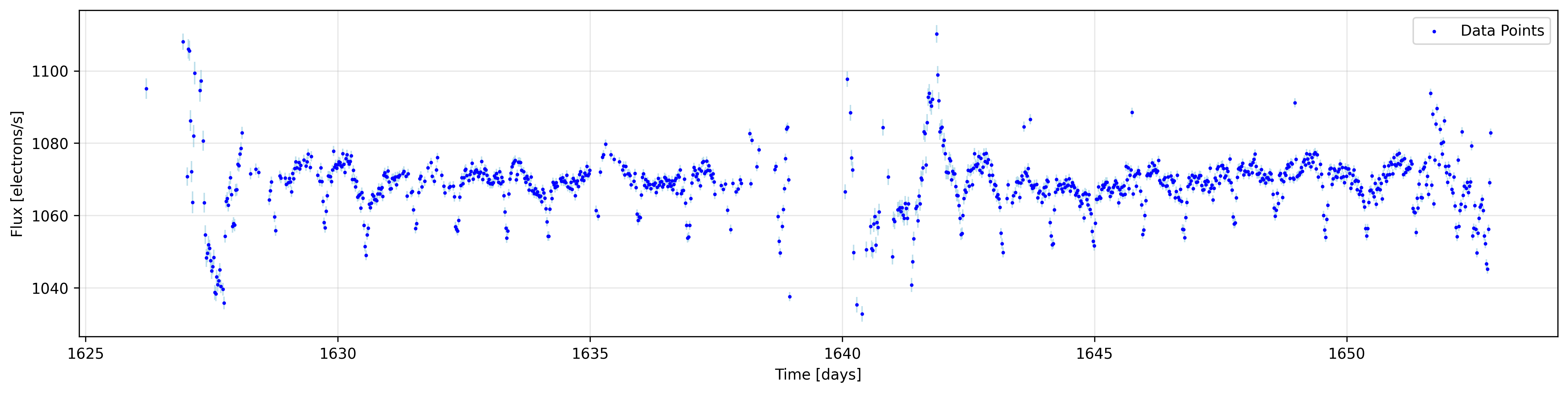}}
\caption{\tes (sector 12) light curve of J1716$-$36
\label{f:tess_j1716}
}
\end{figure*}

\subsubsection{Time-resolved spectroscopy}
Using the MDM 2.4m telescope time-resolved spectroscopy was performed with the aim to determine the binary orbital period of the candidates. Some of the objects are spectroscopic double-lined binaries with emission lines from the accreting gaseous matter and absorption lines from the donor star.  These observations revealed unique periods for 10 out of 11  objects that were observed (J0426$-$25, J0503$-$24, J0503$+$20, J0509$-$03, J0531$-$01, J0559$+$11, J0610$+$31, J0715$+$08, J0726$-$10, J1113$-$17). An attempt to determine the period for the eleventh object, J0729$+$09, remained unsuccessful although a relatively large number of 24 spectra were obtained. The periods found range from 1.6 h (J0503$+$20, DN) to 10.04 h (J0726$-$10, DN). The results for J026$-$10, a NL CV, are shown in Fig.~\ref{f:0726specrv}. For five of these independent confirmation of the period was found from the analysis of \tes data. The detailed analysis of the time resolved data on the polars and NL observed from MDM will be presented separately (Thorstensen et al., in preparation).

\subsection{Diagnostics based on cataloged fluxes}\label{s:catval}
We also use diagnostic diagrams as presented by \cite{schwope+24b} to classify our new objects. We use an X-ray/optical color-color diagram  (CCD) combining \ero and \gai photometry, the \gai-based color-magnitude diagram (CMD), the luminosity-distance diagram, again combining \ero and \gai data, and the \gai variability diagram.
Every object with $L_{\rm X}>10^{32}$\,\lx was thus regarded as an IP candidate, every object with an X-ray to optical flux ratio below $f_{\rm X}/f_{\rm opt} < 1$, that is also located close to the main sequence in the CMD is a good NL candidate, and any object close to the WD sequence is a good candidate for being a low mass-transfer object like a DN (possible WZ subtype) or a polar in its low state.  

The whole suite of all standard diagnostics is shown for one example CV in Fig.~\ref{f:1diag}. Such a profile sheet was generated for each object and used for subclass determination. Not included in that figure are time resolved data obtained by us and its analysis and the \tes light curves. These and the Lomb-Scargle periodograms were used in addition for subclass determination, depending on their availability.

\begin{figure}
\resizebox{\hsize}{!}{\includegraphics[clip=]{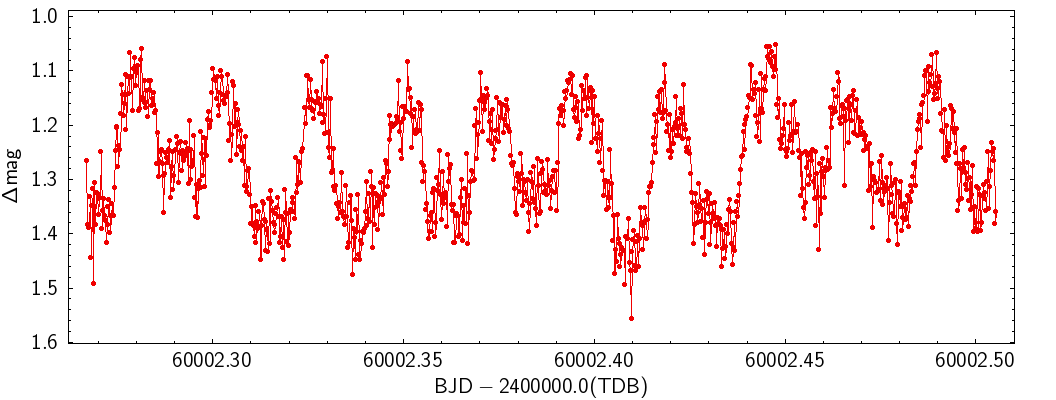}}
\caption{SAAO light curve of J0755$-$17, unveiling the 33\,min spin period of the IP candidate
\label{f:j0755}
}
\end{figure}

\subsection{Analysis of time-resolved photometry}

\tes data were analyzed by computing Lomb-Scargle periodograms. If one or several significant ($>5\sigma$) periods were found, a decision had to be made about its likely origin in the binary. The decision was based on the shape of the power spectrum answering the following questions: Was the period found unique or where many competing periods found? Did the folded lightcurve using the most likely period reveal a shape, that convincingly show orbital phase-dependent variations? Is the period within the range of known orbital periods (between about 80 min and 2 days)?  Our assessment revealed likely orbital periods for 16 of the observed 24 CVs that were observed with \tes and secondary periods for three objects. Secondary periods were regarded as those lying outside the acceptable range of $P_{\rm orb}$. For two objects, J0800$-$43 and J1515$-$-53, these secondary periods were identified with the spin periods of the WDs in IPs, for two other objects, J1302$-$41 and J0531$-$01, slightly discrepant spectroscopic and photometric periods could be associated with superhump periods in the binaries. In J1302$-$41, the secondary period of 96\,min was identified with the likely positive superhump of a DN with an orbital (primary) period of 93 min.  

We classified one object based on comprehensive SAAO photometry alone, without confirmation by a spectrum. It is object J0133$-$81, which shows very deep eclipses. The shape of the eclipse light curve in particular with a first steep ingress (WD ingress) followed by slower ingress \citep[occultation of the accretion stream, like in e.g.~V808 Aur][]{schwope+15}, and the overall shape outside of the eclipse with strong cyclotron beaming and flickering leave no doubt, that this object is a polar (Fig.~\ref{f:lc_j01333}). It was reported previously as a polar in vsnet-alert 27667 but the eclipse was not found in the ATLAS data analysed there. Linear regression of mid-eclipse times spread over 1025 orbital cycles revealed the linear ephemeris of BJD(TDB) $= 2460286.64514(23)+ E \times 0.055018(15)$.

Archival photometry from the ZTF, the CRTS, and ATLAS was mainly used to search for dwarf nova outbursts (DNO), but a period search was also performed where this appeared promising. This revealed unique results in a few cases, a good example is that of the eclipsing DN J0715$+$08 with a period of 5.065508(8) hours which was derived from all the ATLAS data alone (Fig.~\ref{f:asn0715}. 

An interesting catch was that of J1716$-$36. Its \tes light curve, although obtained only with the 30 min cadence, revealed eclipses separated by almost one day (see Fig.~\ref{f:tess_j1716}). The period analysis revealed the extraordinarily long period of 21.78 hours.

On the other side of the period distribution, photometric monitoring of J0047$-$48 with SAAO facilities revealed a unique period of 79.2262$\pm$0.0030 min, which is compatible with the canonical CV minimum period of $82.4\pm 0.7$\,min \citep{knigge+11} only at 3$\sigma$, but better compatible with the revised minimum period of $79.6 \pm 0.2$ min given by \cite{mcallister+19}.

To summarize, the photometric and spectroscopic follow-up of our new CVs revealed stable periodic signals for 44 objects.  Forty of the periods found were regarded as orbital periods. Such an identification is secure for the seven eclipsing objects (J0133$-$81, J0715$+$08, J0742$-$24, J1231$-$42, J1241$-$60, J2143$-$57, J2307$-$40) and for double-lined binaries where the radial velocity could be traced through an orbital cycle (J0426$-$25, J0509$-$03, J0531$-$01, J0726$-$10, J1113$-$17). In eight objects a secondary period could be identified. In two of those objects (J1301$-$41, J0531$-$01) this was likely identified with a positive superhump in a DN, in the other six with the spin period of the WD. One example is J0755$-$17, its $r$-band light curve is shown in Fig.~\ref{f:j0755}, which shows clear periodic modulation of the brightness with $P=33.5$ min and an amplitude of about 0.2 mag. These six were classified as IPs, all have X-ray luminosities in excess of $10^{32}$\,\lx, two of these also have their orbital periods measured, both at 5.4 hours (J0800$-$43, J1515$-$52). The orbital periods of the other four strong IP candidates with $P_{\rm spin}$ tentatively measured could not yet be determined. All the found periods are reported in the online publication of the results table, an abridged version is given in Table~\ref{t:cvs}.

\begin{figure}
\resizebox{\hsize}{!}{\includegraphics[clip=]{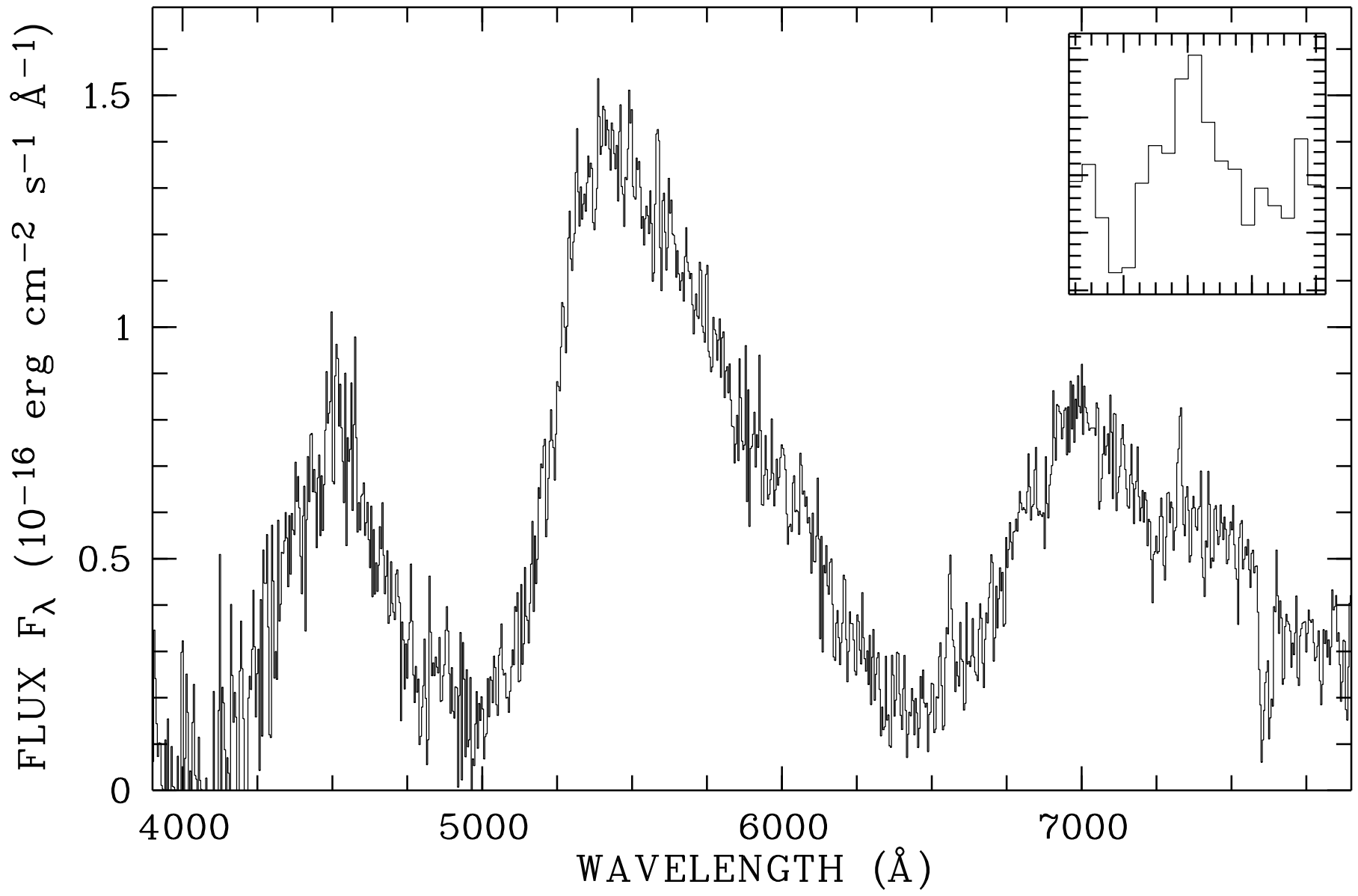}}
\caption{Discovery spectrum of J1754$-$44, dominated by cyclotron harmonic emission from a magnetized plasma with $B=47$\,MG, obtained with EFOSC2 at the NTT. The inset shows the region around the H$\alpha$ line.\label{f:j1754_cyc}
}
\end{figure}

\subsection{Magnetic fields inferred from cyclotron and Zeeman features}
The MCVs may show features in their optical spectra that can be directly used to infer the magnetic field of the WD. These are Zeeman-split and -shifted absorption lines from the WD photosphere or an accretion halo and cyclotron harmonic lines. For the analysis of such features the reader is referred to the prominent examples of MR Ser or V834 Cen, well-studied bright polars \citep{schwope+93}. We found five systems that show either Zeeman or cyclotron lines. 

A particularly spectacular example for cyclotron lines in one of our new CVs is that of J1754$-$44 (see Fig.~\ref{f:j1754_cyc}). The position of the cyclotron harmonics and their width indicate a moderately low plasma temperature of $kT \lesssim 5$\,keV and a field of 47\,MG. There is a second object showing a cyclotron harmonic line, J1241$-$60, an eclipsing polar with $P_{\rm orb}=2.22$\, h orbital period, but the observation of only a single harmonic centered at 5415\,\AA prevents us from measuring a field strength. There is a slight flux enhancement at around H$\gamma$. If this indicates the next higher harmonic, the field would also be of order $45 - 48$\,MG.  

Objects with an indication of Zeeman features are J0507$-$09, J1448$-$48, and J1745$-$50. The first two were classified as polars in their low accretion state, hence their Zeeman features, which appear most pronounced through the $\sigma^\pm$ lines of H$\alpha$, are of photospheric origin. Their centroids indicate a flux-weighted, mean field strength at the corresponding orbital phases, when the spectra were obtained, of $6-7$\,MG and $19-20$\,MG for J0507$-$09 and J1448$-$48, respectively. The field strength at the pole will be higher but is difficult to infer without phase-resolved data. The Zeeman lines of J1745$-$50 with a field strength of $15-16$\,MG were detected in a high-state spectrum, which makes an origin in an accretion halo surrounding the accretion column more likely than a photospheric origin \citep[see e.g.][]{schwope+90}.

\begin{figure*}
\resizebox{0.5\hsize}{!}{\includegraphics[clip=]{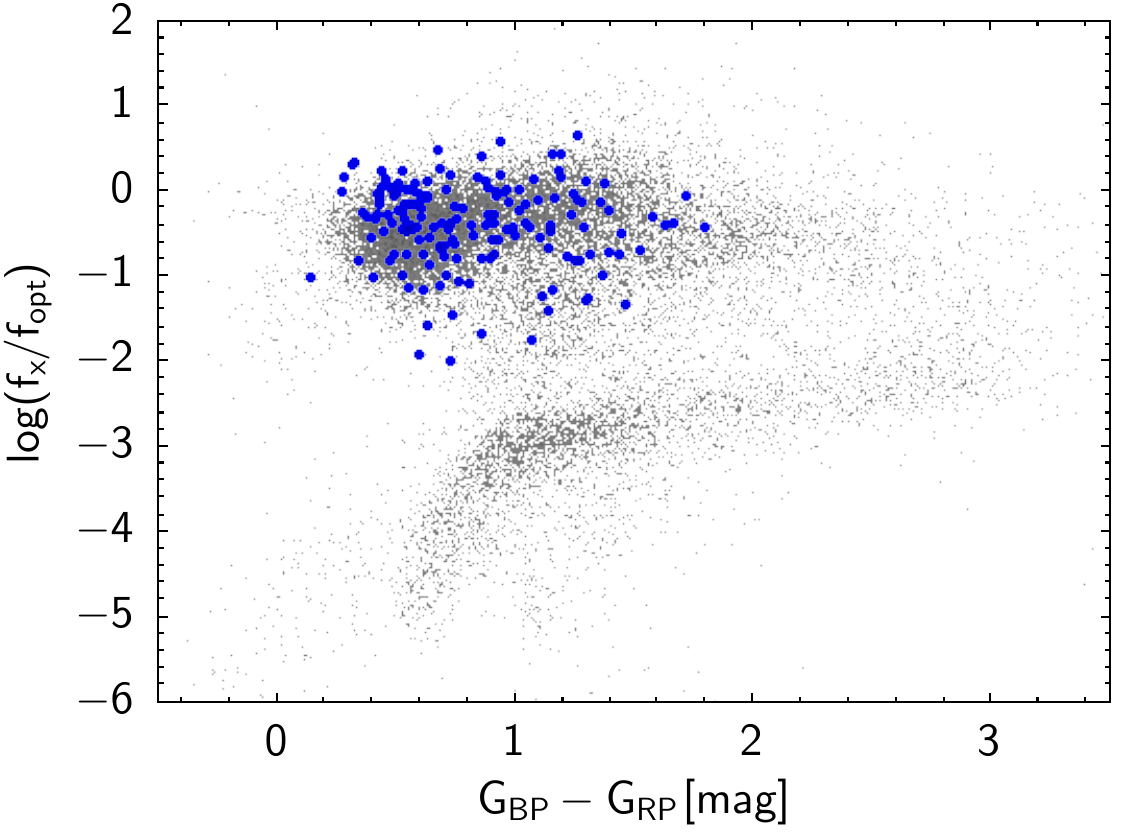}}
\resizebox{0.5\hsize}{!}{\includegraphics[clip=]{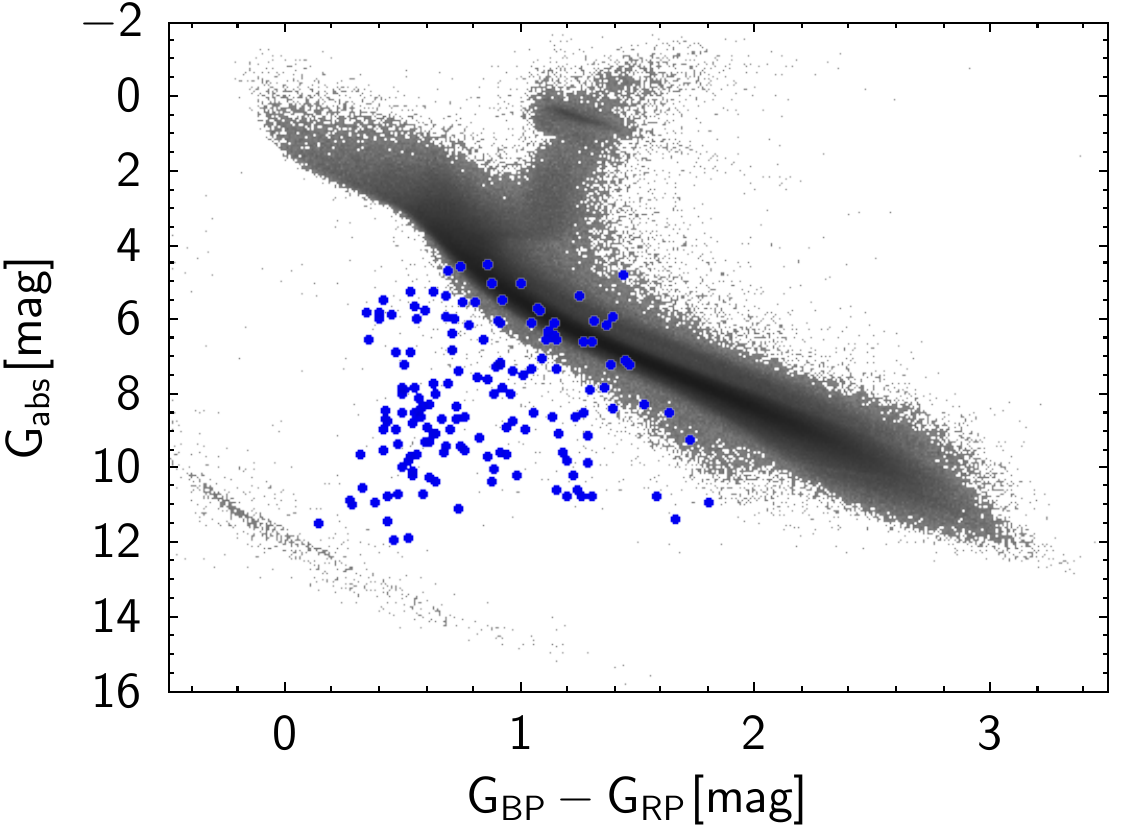}}
\resizebox{0.5\hsize}{!}{\includegraphics[clip=]{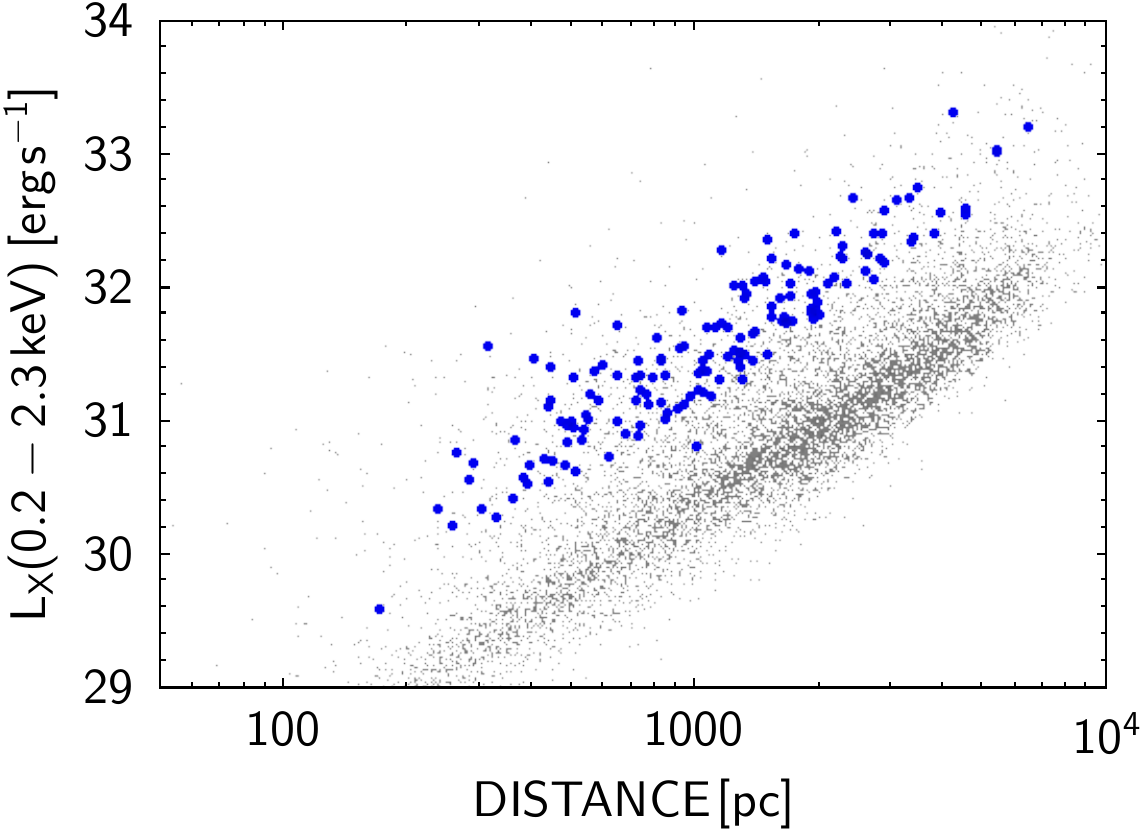}}
\resizebox{0.5\hsize}{!}{\includegraphics[clip=]{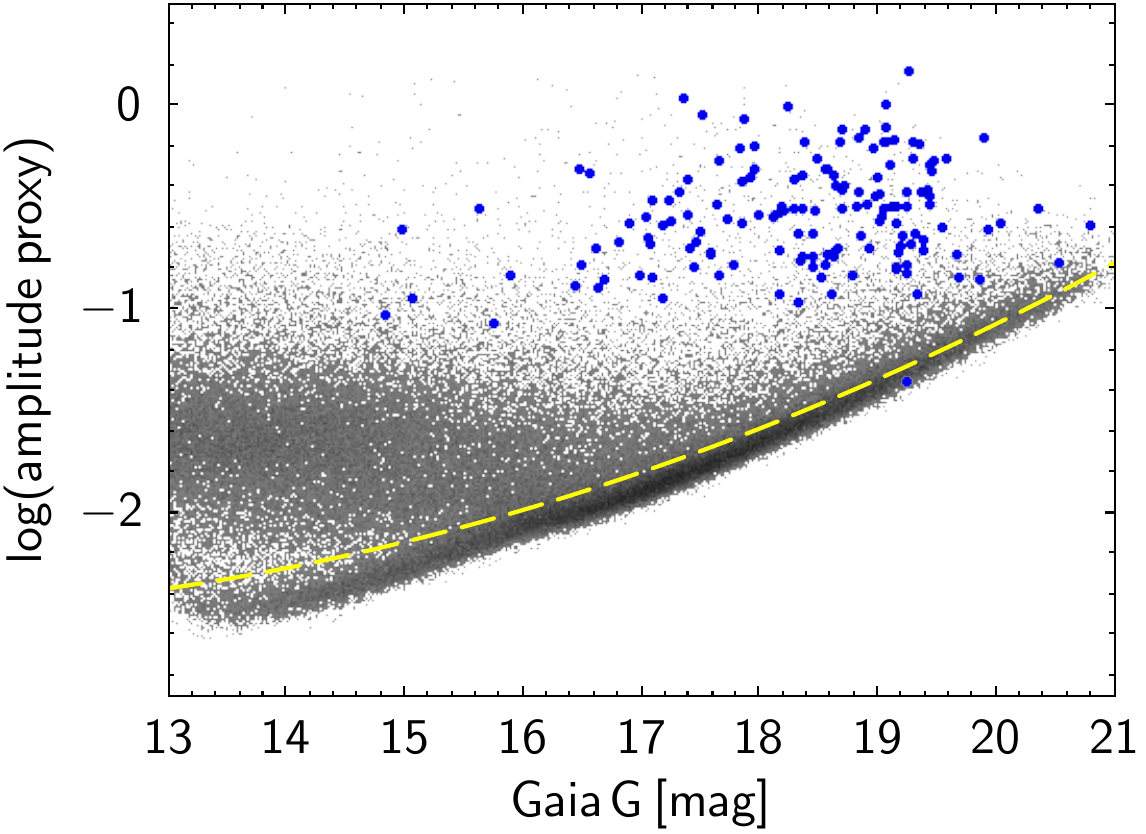}}
\caption{Diagnostic diagrams showing the location of all new CVs compared to unrelated objects or candidate CVs. The CCD is shown upper left, the CMD upper right, the luminosity-distance diagram lower left, and the \gai variability diagram lower right. The construction  of the various diagrams is explained in \cite{schwope+24b}.
\label{f:diag}
}
\end{figure*}

\subsection{Subtype and subclass assignment}
The joint view on the various derived parameters and diagnostic diagrams often gives a good clue on the likely subtype of an object. Examples are: if a periodic modulation of the optical brightness was found with a period below one hour, the object shows the \het emission line and an X-ray luminosity $L_X > 10^{32}$\,\lx it was classified as an IP. If no periodic pulsations were found, the object was classified as an IP candidate. The candidate status is documented in the 'comments' column of Table~\ref{t:cvs}. An object was classified as an MCV, if showing \het with a line luminosity separating it from the NL (see right panel of Fig.~\ref{f:lines}), while showing a moderate $L_X$. A polar was classified as such if its spectrum showed \het emission, and one unique period was found, identified as \porb. The shape of the continuum, e.g.~showing a cyclotron radiation or Zeeman absorption, was also taken into account and could lead to a unique classification, even if a period determination was not possible. A DN was identified as such if the long-term light curves showed DNO, if the emission lines showed the typical double-peaked profile from the accretion disk, the line luminosity was low and the line ratios separating it from the MCV range. An NL classification was assigned, if a luminous bright blue continuum was found in the optical, the absolute magnitude high and the X-ray to optical flux ratio comparatively low. The VY Scl (VY) subclass was assigned if the long-term light curve shows extended phases of reduced brightness. Sometimes the identification was seemingly unique, sometimes not. In the latter case, an object was tentatively classified and was labeled as 'cand' in the comments column of Table~\ref{t:cvs}. While running the classification exercise we were sometimes left with too sparse information and were not able to assign a subtype. Such objects were assigned the general category as 'CV'.

\section{Results}\label{s:res}
\subsection{Identification summary}
Our list of newly identified CVs from eRASS1 comprises 156 objects. All but one were confirmed as CVs through low-resolution spectroscopy. The one object without a spectrum was identified through its unique optical light curve showing an eclipse by the donor star and the typical modulation of its light curve by an irradiated accretion stream and by cyclotron beaming (See Fig.\ref{f:lc_j01333}). The list contains two objects that were meanwhile (after taking the original discovery spectrum and while working on this publication) identified by others and published separately with our participation, namely the polar J1527$-$20 \citep{ok+24} and the white dwarf pulsar J1912$-$44 \citep{pelisoli+23, schwope+23}.

\begin{table}[t]
\begin{center}
\caption{Identification summary: CV subtypes and subclasses\label{t:subty}}
\begin{tabular}{ll|rr|rr}
\hline \hline
Subtype & Subclass & \# & [\%] & \# & [\%]\\
&&\multicolumn{2}{|c}{Total} & \multicolumn{2}{|c}{$<1$\,kpc} \\
\hline
NL &  all & 19& 12 &  4 & 5\\
   &   VY & 2 &  1 &  1 & 1\\
DN &  all & 67& 43 & 37 & 47\\
     & WZ & 2 &  1 &  2 & 3 \\
MCV & all & 63& 40 & 24 & 30\\
    &  IP & 13 & 8 &  0 & 0\\
  & Polar & 20 &13 & 14 & 18\\
      CV & & 6 & 4 &  1 & 1\\
WD pulsar& & 1 & $<1$& 1 & 1\\
\hline 
Total & & 158 & 100 & 79 & 100\\
\hline
\end{tabular}
\end{center}
Notes: Percentages were rounded to the next integer. 
\end{table}

The classification summary is given in Table~\ref{t:cvs} with one line per object. It lists the IAUNAME from \ero DR1, the \gai ID of the counterpart, the subtype and subclass, a very short comment on the object, and the main and secondary periods found. The full version of the table is published online (see Sect.\ref{s:data}). Column formats are described in Tab.~\ref{t:cat_col}.

The full identification table lists combined catalog values from \ero DR1 (IAUNAME, DETUID, X-ray coordinates, number of detected photons, ML\_CTS, and derived flux - ML\_FLX, respectively) and from \gai DR3 (coordinates, $G$-band magnitude and color $BP-RP$, parallax), derived quantities (distance, r\_geo from \cite{bailer-jones+21}, absolute $G$ magnitude, $\log L_X$ ($0.2-2.3$\,keV), $\log f_X/f_{\rm opt}$), the CV subtype and a brief comment on the CV plus the measured periods. Most of the subtype classifications are, based on the current evidence, regarded as preliminary. If a tentative classification was possible but remained ambiguous, we labeled the object as 'cand'.  The 'comment' column also holds information about the subclass, eclipses, magnetic fields and previous publications. 

Taking the preliminary subclassification at face value, the breakdown into subtypes as given in Table~\ref{t:subty} emerges. This has about equal fractions of MCVs (with a large sub-population of IPs) and DN, both at $\gtrsim$40\%. It also has a 12\% fraction of NLs and a small fraction of non-classified objects. The identification program was undertaken with the primary aim to validate our procedure to select likely CVs as optical counterparts of newly found \ero sources and by these means to get a preview of the full harvest from a complete identification program, which would allow to compose volume-limited samples of different sub-categories. Our survey is flux-limited but it is not complete, hence is biased in several ways, which is reflected in the breakdown of subclasses. X-ray luminous objects will be overrepresented in the sample. In the same way, a certain bias will exist towards bright optical objects through the preference of observers sitting at the telescope when insufficient telescope time was available to complete the whole identification program (there were more targets available than could be observed given the limits on telescope time). Such biases become obvious if we look at the breakdown of objects within a reasonable distance limit. Within 1\,kpc we get a lower fraction of MCVs, actually no new IP lies within 1\,kpc, we get less NLs (5\%) and a higher fraction of the less luminous DN, which make up almost 50\% of the more local (1 kpc vs unconstrained) sample. 

\subsection{The new \ero eRASS1 CVs in the CV landscape\label{s:landsc}}
Fig.~\ref{f:diag} gives an overview of the main properties of the new CVs in four diagnostic diagrams, a CCD, a CMD, an X-ray luminosity-distance diagram, and an optical variability diagram using \gai data. Each of the diagrams shows the new CVs with blue symbols on a suitably chosen background (either unrelated objects with \gai and/or \ero data or other CV candidates, all background objects are shown with gray symbols). The chosen background serves as a guide for the eye and the layout of the graphs. The background objects shown are the same as in \cite{schwope+24b}. The new CVs cover a wide range in apparent and absolute optical brightness ($G = 14.8 - 21.8, M_G = 4.5 - 11.9$), and several orders of magnitude of the X-ray luminosity, $\log L_{\rm X} \mbox{\rm  (erg/s)} = 30.1 - 33.3$, and distances between 170\,pc and several kpc. All CVs have a X-ray to optical flux ratio $\log (f_{\rm X}/f_{\rm opt}) > -2$, all are bluer than $B-R=2$ and all were found to be strongly variable. The location of the new objects in the diagnostic diagrams reflects the location of the objects in the training sample. Because of the applied flux cut, $f_x> 1 \times 10^{-13}$\,\fergs, the sample presented here is sparsely populated in the region where the white dwarf-dominated, low accretion-rate objects are typically encountered, close to the WD sequence in the CMD \citep[period bouncing objects, WZ Sge objects, and LARPS][]{munoz-giraldo+24,schwope+02b} and Hernandez-Diaz et al.~(2026, subm.).

In the CCD, all new CVs are located away from the stellar branch (coronal emitters), defined by the arc stretching from $(B-R;\log(f_x/f_{\rm opt}))=(0.6;-5.5)$ to $(3;-2.5)$, but are found almost exclusively on top of the cloud of AGN at $\log(f_x/f_{\rm opt})) \simeq -0.5$. There is an excursion towards lower flux ratios for the NLs with their hot, hence bright accretion disks. Perhaps not too surprisingly given the name as cataclysmic {\it variables}, all new CVs are found to be \gai variable, that means well separated from the yellow non-variability line in the lower right panel of Fig.~\ref{f:diag}. The \gai collaboration has generated and published with DR3 many lists of variable objects, among them a list comprising 7306 objects of CV candidates \citep{eyer+23}. Interestingly, only 53 of our objects are found in their list of CV candidates, more than 100 new CVs escaped their tentative classification which was solely based on \gai variability.

The luminosity-distance diagram (Fig.~\ref{f:diag}, lower left panel) uses as background objects the CV candidates that are proposed for spectroscopic follow-up in SDSS-V \citep[see][their Fig.~11]{kollmeier+26}. SDSS target selection is based on the stack of the first three eRASS with a corresponding lower sensitivity limit (\ero DR2; Ramos-Ceja et al., submitted to A\&A). The minimum X-ray flux chosen for the sample presented here is about a factor 10 above the sensitivity limit of DR2. This explains the large separation of our sample from the main sample of  objects foreseen for spectroscopic follow-up in SDSS-V. The sample compiled here clearly represents the bright tail of the distribution. The number of X-ray photons collected in \ero DR1 for our identified CVs varies between 6 and 472 with a median of 38. Even at the lowest number of photons, the X-ray detection is secure with a detection likelihood {\tt det\_{ml} > 13}\footnote{The DR1 catalog lists objects down to {\tt det\_{ml} = 6}}. For a definition see \cite{brunner+22}. 

\begin{table}[b]
\begin{center}
\caption{X-ray detection of the new DR1 CVs presented here in the various \ero surveys\label{t:xdet}}
\begin{tabular}{lrr}
\hline \hline
Survey & \# detections & [\%]\\
\hline
eRASS1 & 156 & 100\\
eRASS2 & 148 & 95\\
eRASS3 & 137 & 88\\
eRASS4 & 145 & 93\\
eRASS1 \&\& 2 \&\& 3 \&\& 4 & 129 & 83\\
eRASS:3 & 156 & 100 \\
eRASS:5 & 156 & 100\\
\hline 
\end{tabular}
\end{center}
Notes: Percentages were rounded to the next integer. The 4$^{\rm th}$ row gives the number and faction of CVs that were found in each eRASS, the last two rows give the numbers for the stacks of the first three \cite[published as \ero DR2][]{ramos-ceja+26}, and of all \ero surveys (unpublished).
\end{table}

The comparison of the new DR1 CVs and their detection or non-detection in other \ero catalogs is instructive because it gives some further insight into the sample properties. We searched for X-ray detections of point sources within 30 arcsec of the \gai positions of the new CVs in the X-ray catalogs of eRASS2, eRASS3, eRASS4, the stack of the first three eRASS \cite[called eRASS:3, published as \ero DR2][]{ramos-ceja+26}, and the stacked catalog based on all survey data, called eRASS:5. The latter is a preliminary version of the catalog that will be published as DR3. The results are listed in Tab.\ref{t:xdet}. All individual surveys have about the same flux limit of $4\times 10^{-14}$\,\fergs \citep{merloni+24}. This is clearly above the chosen limit of $10^{-13}$\,\fergs for the selection of the candidate counterparts for this study. Hence, the variable source content among the surveys is mainly due to CV variability, not missing sensitivity or other technical limitations. A fraction of about 10\% of the sources was dimmed by considerable factors between 3 and 17 when comparing the eRASS1 and the eRASS:5 flux. This explains why they could not be discovered in any of the following surveys after eRASS1. Of course, other sources will have been brightened by similar factors, so that the source content in a snapshot survey like ours is subject to considerable changes. Such effects need to be taken into account when addressing the completeness of the X-ray survey and when implications are drawn for the space density and the derived CV luminosity functions. 
\begin{figure}
\resizebox{\hsize}{!}{\includegraphics[clip=]{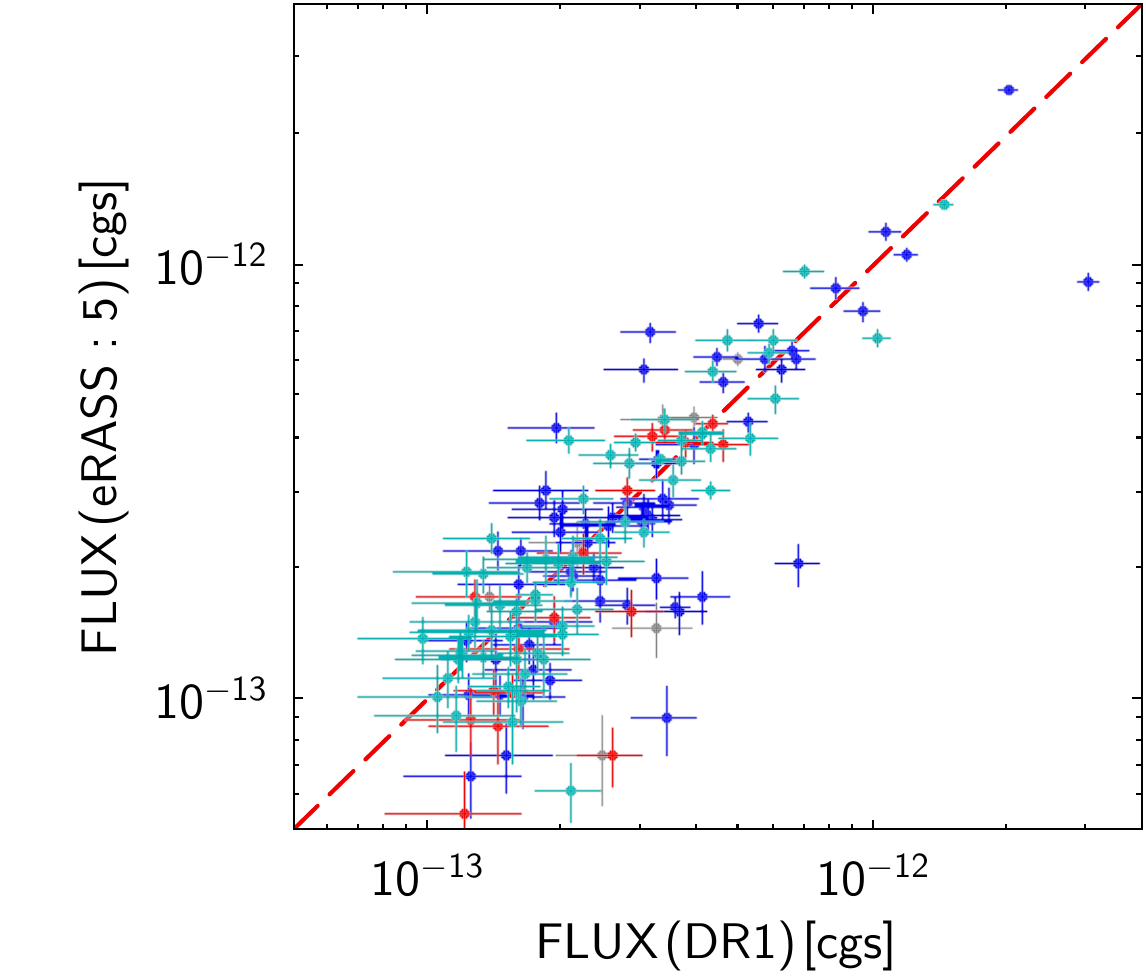}}
\caption{X-ray variability of new CVs. Non-classified CVs are shown with grey symbols, all sorts of MCVs with blue, NL with red and DN/WZ with cyan symbols.
\label{f:cvser5}
}
\end{figure}

All new CVs from our eRASS1 study are also found in the eRASS:3 and the eRASS:5 catalogs, in the latter at a median separation of 1.4 arcsec. In eRASS:5 we find the CVs with a minimum of 11, a maximum of 2580, and a median of 131 photons. An X-ray spectral and timing analysis of the new CVs based on all \ero survey data will be published separately. A comparison of the mean X-ray fluxes in eRASS1 and eRASS:5 is shown in Fig.~\ref{f:cvser5}. The X-ray flux ratio $f_{5} / f_{\rm DR1}$ varies between 0.26 and 2.22 with a mean of $0.96\pm0.35$. Most CVs show a moderate variability pattern, but some CVs were fainter in eRASS:5 by large factors. Strong variability (more than 50\%) is found among all source classes, but MCVs/polars represent the largest fraction of about 50\% of those. This is not surprising due to the notorious changes between high and low accretion states in polars and the immediate response in observed properties. On the other hand, in eRASS:5, 14 of the CVs from our sample were found below $f_X = 10^{-13}$\,\fergs but most of these are non-magnetic CVs,
only three were classified as MCVs (J1738$-$34, J1754$-$44, J1904$-$54). The polar with the pronounced cyclotron lines (Fig.~\ref{f:j1754_cyc}) is among those. It appears this system seemed to be encountered in its high state during eRASS1, and it fell into its low state afterwards, and was still in the low state when our identification spectrum was taken (July 2023). 

In Fig.~\ref{f:hr1lx} we show the new CVs in the luminosity -- hardness ratio plane. The hardness ratio is here defined as  HR1$=(H-S/(H+S)$), with $S$ and $H$ being the counts in the soft band between 0.2\,keV and 0.5\,keV, and the hard band between 0.5\,keV and 1.0\,keV, respectively. The chosen hardness ratio is sensitive to a blackbody-like soft X-ray component in addition to the usual thermal spectrum. Such a soft excess was often encountered in MCVs, which show a hard thermal plasma ($>10$\,keV) from the accretion column and a soft blackbody-like spectral component from the accretion-heated area on the white dwarf surrounding the immediate footpoint of the accretion column. HR1$>0.3$ is compatible with just an absorbed optically thin thermal spectrum. A value of HR1$<0.3$ indicates the presence of a soft component. 

We find 14 soft sources covering almost the whole range of X-ray luminosities found among our CVs. All of them were classified as MCV candidates before inspecting their X-ray spectral hardness. One of the soft sources is an eclipsing polar, two were classified as NL/IP candidates. The presence of a soft component suggests these are new IPs, because a fraction of IPs are known to have soft X-ray spectral components. Soft X-ray emission from boundary layers in NLs was expected but never confirmed \citep[see ][for a recent X-ray spectral analysis of a NL and a discussion of its X-ray spectra]{cuneo+26}. Such a finding would be very important. A unique subclassification of the soft sources is needed which could not be given here in any case. It typically requires time-resolved follow-up observations (photometry, spectroscopy) at optical and perhaps also at X-ray wavelengths.

\begin{figure}
\resizebox{\hsize}{!}{\includegraphics[clip=]{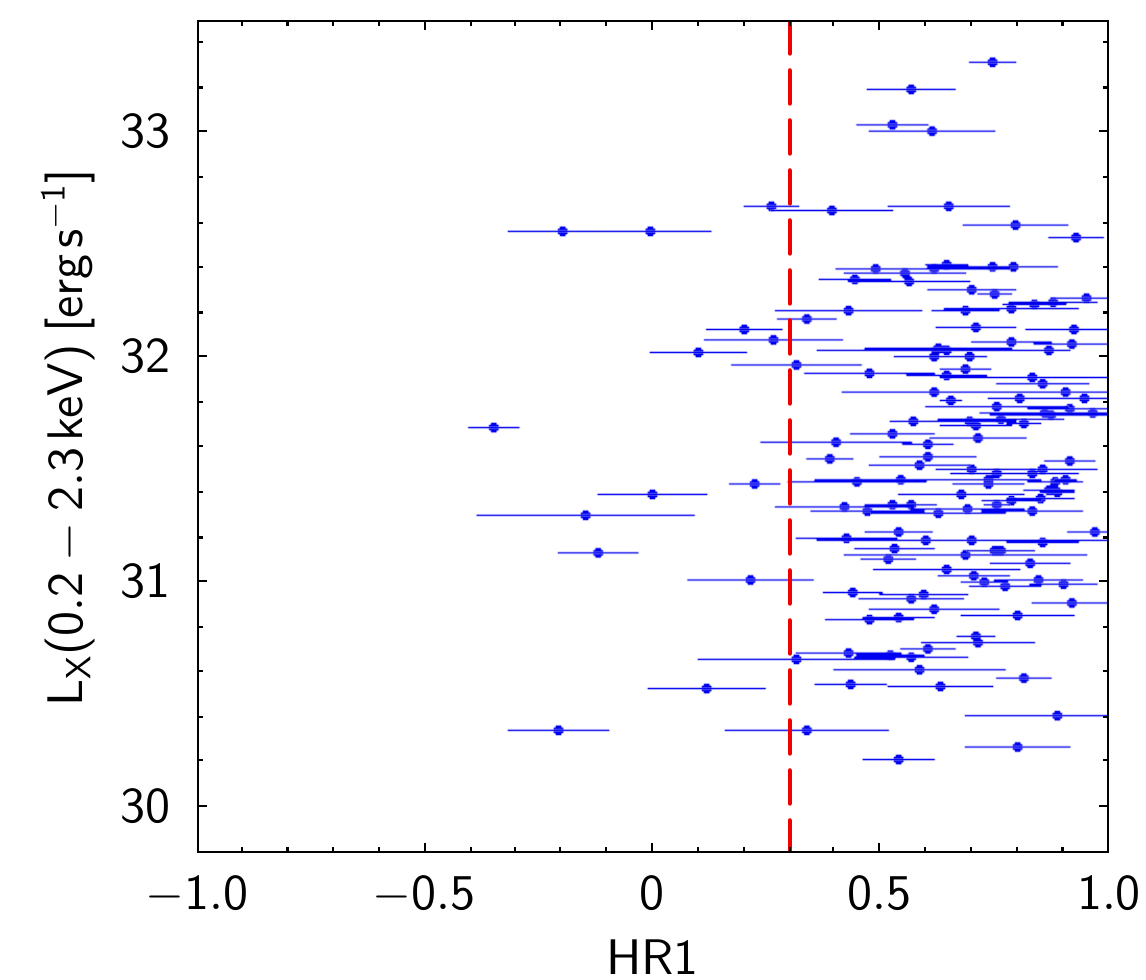}}
\caption{X-ray spectral hardness of new CVs. The red dashed line separates spectra which might be of pure thermal origin (HR1$>$0.3) from those that have an additional soft component. 
\label{f:hr1lx}
}
\end{figure}

\subsection{Identifications of further \ero sources}
Not all spectra that were obtained in the course of our identification program revealed CV features. We are confident that a further five objects observed by us are the correct counterparts of the corresponding eRASS X-ray sources, because their spectra revealed their nature as AGN. One further is a CV candidate, but no spectrum could be taken due to the lack of telescope time, just a spectrum of a nearby candidate which turned out to be an inactive M-star, which therefore seems to be unrelated to the X-ray source. 

The combined optical and X-ray properties of these other objects from the \ero and \gai catalogs were CV-like, in particular the AGN were regarded to belong to our own galaxy based on their inferred distances from \cite{bailer-jones+21} . We determined their redshifts from their emission lines and list those in Table \ref{t:oxid}. Their locations in the diagnostic diagrams, the same layout as used above for the CVs (Fig.~\ref{f:diag}), is shown in Fig.~\ref{f:oxid}. From there it is apparent why the objects were considered CVs, their cataloged parameters are very similar to those of the CVs.

\section{Discussion and conclusion\label{s:disc}}

We have performed exploratory spectroscopic follow-up observations of CV candidates selected from the first \ero all-sky survey (eRASS1). All X-ray sources from the preliminary source catalog that was used at the time when targets were selected for spectroscopic follow-up are also contained in \ero DR1 and got a proper IAU name (see Tab.~\ref{t:cvs}). Targets were selected through a probabilistic match between point-like X-ray sources and objects listed in \gai DR3. Only X-ray sources with X-ray flux $f_x (0.2-2.3 \mbox{keV}) > 10^{-13}$\,\fergs and matching pairs with a likelihood larger than 50\% were chosen for follow-up. Known objects at the time of target selection in January 2021 were discarded from our target lists. Spectroscopy using facilities in South and North America (ESO/NTT and Kitt Peak/MDM), South Africa (SALT and other SAAO telescopes), and La Palma (NOT) was performed to securely identify the chosen candidates. We had more targets than we could observe, thus our flux-limited mini-survey gives a sneak preview of the full harvest after spectroscopic identification programs of \ero sources are completed. The analysis of our spectroscopic identification program was complemented by dedicated time-resolved photometric observations from Northern and Southern facilities (MDM, SPECULOOS, SAAO) and by archival photometry from the ZTF, the CRTS, the ATLAS and the ASAS-SN surveys as well as from TESS. 

We identified 156 CVs, most of them new to the literature. A few objects were meanwhile (after the start of our identification program) reported as CVs \citep{inight+25,szkody+21, ok+24}, but their detection as \ero X-ray sources is not reported elsewhere and they are kept here for this reason. Several other objects confirmed here as CVs were known to the literature but mis-classified as WD \citep{gentile_fusillo+19, gentile_fusillo+21, vincent+24}, as subdwarf \citep{culpan+22}, or as eclipsing binary \citep{mowlavi+23}, because spectral information was missing.  We identified a further eight \ero sources as either coronal emitters or AGN. The new CVs cover a wide range in X-ray luminosity, in absolute magnitude, and distance. We found almost all subtypes of CVs, but a selection bias towards bright, luminous objects is present in our sample because of the applied flux cut. Twenty-four (13\%) of the new CVs are located within 500\,pc. It is the volume with this radius where one may expect a 90\% completeness of CVs through spectroscopic follow-up of \ero sources \citep{schwope+24b}. Twelve objects in our sample are rather X-ray luminous with  $\log L_{\rm X} \mathrm{(erg\, s}^{-1})> 32.5$. These are prime candidates for IPs, a potential substantial increase in the general population, and the first few could be confirmed by us through optical time-resolved photometry and are reported here.

In the optical, we found a periodic signal for 44 of our new objects. Mostly this could be interpreted as the orbital period. In some cases we found a period below 60 minutes that was interpreted as spin period in an IP candidate.  The orbital periods range from 79.23 min to almost 22 hours. We found seven eclipsing systems, which is a lower limit to the true number due to missing time-resolved photometry. In five objects we found magnetic features in their optical spectra, which are either photospheric or halo Zeeman lines or cyclotron lines from the accretion plasma. 

We could tentatively assign a CV subtype to most of our new discoveries using the broad categories DN (43\%, with subclass WZ), NL (19\%, with subclass VY), MCV (41\%, with subclasses IP and polar), WD pulsar. A few objects (4\%) remained without assigned subtype; they are simply labeled as 'CV'. However, many of the assigned subtypes are left with the 'candidate' tag, their full specification requires dedicated time-resolved follow-up observations at X-ray and/or optical wavelengths, which is already initiated. The properties of the sample as established in this work is a rich resource to select objects for detailed follow-up studies, to determine e.g.~eclipse profiles, to address time variability, or disentangle their X-ray spectra, to name just a few.

\subsection{Selecting optical counterparts to \ero X-ray sources}

In the following paragraphs we discuss the results of our target selection compared to other routes, that were described in the literature. \cite{freund+24} present a method to identify the coronal content of eRASS1. They found 149,311 stellar counterparts to eRASS point sources, all of them with assigned \gai DR3 properties. We matched their catalog with our list using the given GaiaID and found 38 matching pairs. Under ideal conditions, the combined X-ray to optical flux would cleanly separate the two source classes. Therefore, the number of matches appears high. The matches have no preference for CV subtype, even two IPs are among the matching pairs. The X-ray luminosity ranges from $\log L_x (\mbox{erg\,s}^{-1}) = 30.0$ to $32.7$. The probability $p_{\rm coronal}$ given by \cite{freund+24} ranges from 0.55 to 0.99, and is evenly distributed among those extremes. This flat distribution for the CVs in the stellar sample is different from the large body of their \gai counterparts which shows a steep increase towards $p_{\rm coronal} = 1$, hence the contaminating CVs showed a slightly awkward behavior in their distribution. The few CVs (38 objects) are a minor contamination for the \cite{freund+24} coronal sample. On the other hand, if we had removed the objects listed by them as likely stars under the assumption that their identification would be correct, this would have severely reduced our success rate. 

\cite{salvato+25} published a comprehensive list of optical counterparts to eRASS1 X-ray point sources that were identified using Gaia DR3, CatWISE2020, and Legacy Survey DR10 (LS10) with the Bayesian NWAY algorithm and trained priors. Sources were classified as Galactic or extragalactic via a machine-learning model combining optical/IR and X-ray properties, trained on a reference sample. Matching their counterpart list with our CV list using the given \gai DR3 source IDs revealed 149 matches. The nine non-matching sources are J0841$-$26, J0913$-$62, J1053$-$58, J1058$-$56, J1220$-$58, J1236$-$66, J1522$-$65, J1552$-$65, J1552$-$60, and J1636$-$50.
\cite{salvato+25} list several probabilities assigned to each matching source.  
$p_{\rm any}$ gives the probability that for the X-ray source there exists a counterpart in the cross-matched catalog, the lower the value, the more probable it becomes that the matching source is just a chance association. 
The probability $P_{\rm i}$ gives the relative probability between possible counterparts to be the true counterpart. The matching sources between Salvato's counterpart list and our sample have a flat distribution in their $p_{\rm any}$ between 0 and 1, but their $p_{\rm i}$ is persistently larger than 0.6.  The large values found here implies that the matching source was the only one available. Ninety-one of the matching sources have a likelihood $p_{\rm any} > 0.5$ (61\%, good matches), while less than a quarter (36 objects) have $p_{\rm any} > 0.9$ which can be regarded as excellent matches. All five AGN identifications (Tab.~\ref{t:oxid}) are also given as likely counterparts in the list by \cite{salvato+25}. 

\cite{rodriguez+25} describe their method to construct CV samples from eRASS1 and \gai DR3, which they believe to be complete and volume-limited. We tested what the impact of their cuts applied to \gai data is for the CV sample presented here. Their “X-ray main-sequence” criterion alone recovers 152 of the 156 sources, with the remaining six excluded solely by their hard color cut of $BP-RP=1.5$. However, out of the 156 CVs, only 54 satisfy the full set of \cite{rodriguez+25} conditions. Relaxing the \gai astrometric requirements introduces numerous spurious candidates to their sample selection. It would improve completeness at the expense of a significantly higher contamination. Thus, the \cite{rodriguez+25} methodology yields a pure but severely incomplete candidate list. Using it to establish space densities and luminosity functions will yield biased results. 

A significant fraction (53 objects) of the 156 CVs presented here appeared as CV candidates selected from \gai variables in \cite{eyer+23}. Indeed, our variability analysis (see Fig.~\ref{f:diag}, bottom right) revealed strong variability of all our CVs. Hence, the \gai variability catalog of CV candidates is an interesting complement to select CVs within the \gai magnitude range, but it finds only about one third of our objects. Relying on the \gai variability criterion exclusively would miss most of the CVs identified here.

We also tested, if the new CVs are listed in the catalog of WDs by \cite{gentile_fusillo+21}. We found 33 matching \gai DR3 source\_IDs, but only seven of those have $P_{\rm WD} > 0.5$. Hence, most matching WD candidates were regarded as weak candidates which is confirmed through the current work.  

\subsection{Summary and outlook}
To summarize, the combination of \ero X-ray detections with the \gai survey in the way described here provides a unique resource to establish large, volume-limited samples of CVs with an expected low selection bias. Establishing these samples requires comprehensive optical spectroscopic follow-up, targeting also possible counterparts with a low likelihood $p_{\rm CV}$. This will lead to a significant number of contaminants to reach completeness. This is due to the fact, that CVs are a minority fraction among all point-like X-ray sources. Hence, training samples for robust CV selection from \ero/\gai matching pairs are difficult to establish. The sample generated in this work was used to further train the machine-learning algorithm to select targets for spectroscopic follow-up with 4MOST \citep{dejong+19, chiappini+19}, which will provide the ultimate CV sample in the southern hemisphere.

\section*{Data availability}\label{s:data}
The eRASS1 data used here were published in \cite{merloni+24}. \gai data are available through public archives. The raw spectral data obtained for this program from the ESO/NTT are stored in the public ESO archive. Other spectral or photometric data  can be made available upon reasonable request. Table \ref{t:cvs} is available at the CDS via anonymous ftp to cdsarc.cds.unistra.fr (130.79.128.5) or via https://
cdsarc.cds.unistra.fr/viz-bin/cat/J/A+A/vol/page. The graphical products used for screening the individual objects are available on zenodo.org under DOI 10.5281/zenodo.20833453.

\begin{acknowledgements} 
We thank an anonymous referee for constructive criticism that helped improving the quality of the paper. \\ 

This work is based on data from eROSITA, the soft X-ray instrument aboard SRG, a joint Russian-German science mission supported by the Russian Space Agency (Roskosmos), in the interests of the Russian Academy of Sciences represented by its Space Research Institute (IKI), and the Deutsches Zentrum für Luft- und Raumfahrt (DLR). The SRG spacecraft was built by Lavochkin Association (NPOL) and its subcontractors, and is operated by NPOL with support from the Max Planck Institute for Extraterrestrial Physics (MPE). The development and construction of the eROSITA X-ray instrument was led by MPE, with contributions from the Dr. Karl Remeis Observatory Bamberg \& ECAP (FAU Erlangen-Nuernberg), the University of Hamburg Observatory, the Leibniz Institute for Astrophysics Potsdam (AIP), and the Institute for Astronomy and Astrophysics of the University of Tübingen, with the support of DLR and the Max Planck Society. The Argelander Institute for Astronomy of the University of Bonn and the Ludwig Maximilians Universität Munich also participated in the science preparation for eROSITA. The eROSITA data shown here were processed using the eSASS software system developed by the German eROSITA consortium. \\

Based on observations made with ESO Telescopes at the La Silla Paranal Observatory under program IDs 110.24DX, 111.24Q0,  114.27SL, 115.28JL, and 116.29JQ.\\

This work was supported by the research unit FOR2990 (eRO-STEP: Stellar endpoints with eROSITA) funded by the Deutsche Forschungsgemeinschaft.\\

GL, KK, and VC were supported by the Deutsches Zentrum für Luft- und Raumfahrt (DLR) under Contract Nos.~50 QR 2504, 50 OX 2301, and 50 OR 2405. \\

MRS thanks for support from the internal research project PI\_LIR\_26\_07 at USM.\\

This work has made use of data from the Asteroid Terrestrial-impact Last Alert System (ATLAS) project. The Asteroid Terrestrial-impact Last Alert System (ATLAS) project is primarily funded to search for near earth asteroids through NASA grants NN12AR55G, 80NSSC18K0284, and 80NSSC18K1575; byproducts of the NEO search include images and catalogs from the survey area. This work was partially funded by Kepler/K2 grant J1944/80NSSC19K0112 and HST GO-15889, and STFC grants ST/T000198/1 and ST/S006109/1. The ATLAS science products have been made possible through the contributions of the University of Hawaii Institute for Astronomy, the Queen’s University Belfast, the Space Telescope Science Institute, the South African Astronomical Observatory, and The Millennium Institute of Astrophysics (MAS), Chile.

\end{acknowledgements} 

\bibliographystyle{aa}
\bibliography{newdr1cvs}%

\begin{appendix}

\onecolumn
\section{Illustration of graphical products used for identification and subtype determination of eROSITA DR1 CVs.}\label{a:cvprop}

\begin{figure}[h]
\begin{center}\resizebox{0.7\hsize}{!}
{\includegraphics[]{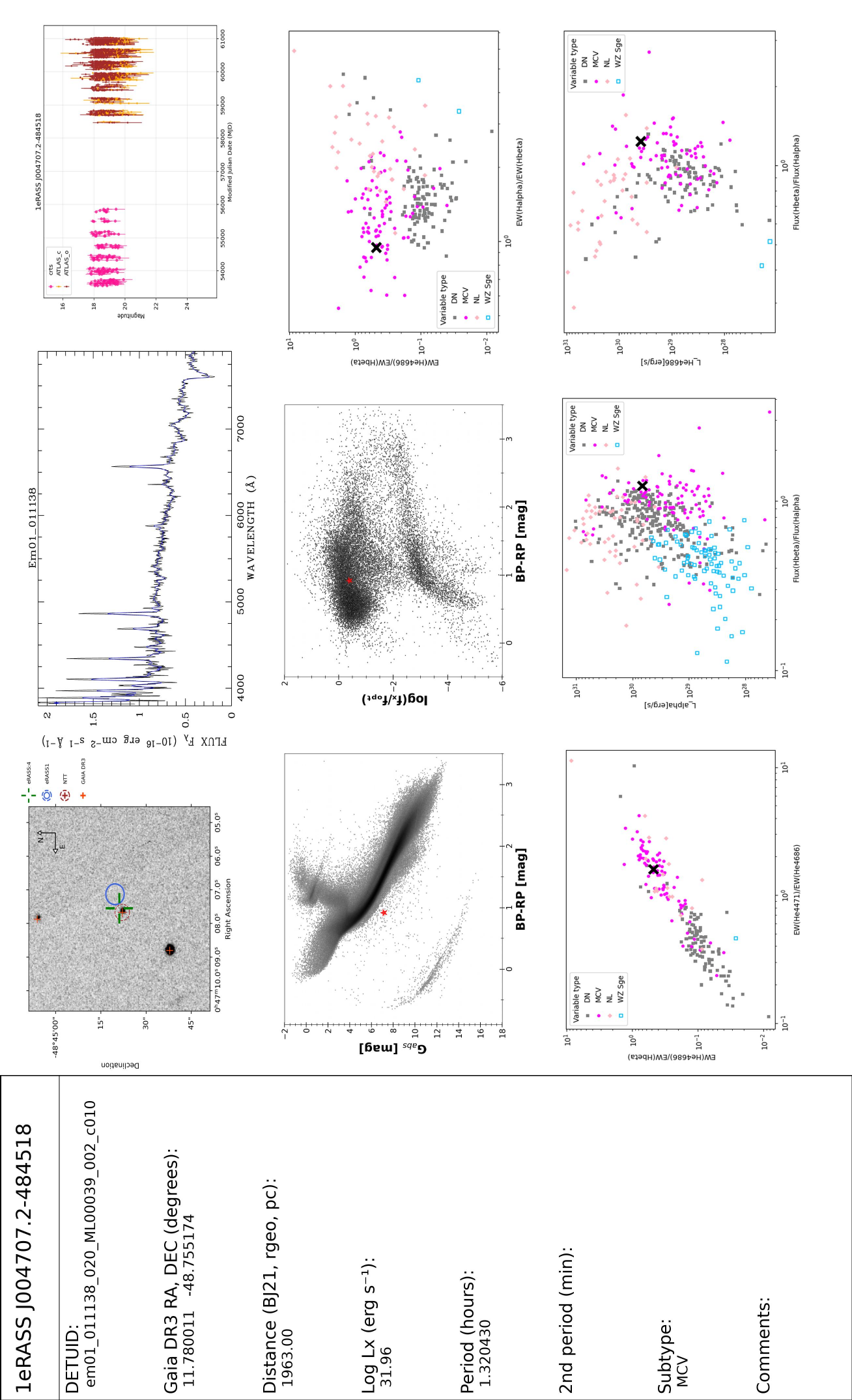}}
    \caption{Suite of standard diagnostic diagrams used for identification and classification (see Sect.~\ref{s:data}), here as one example for object J0047$-$48. All diagrams for all CVs are published online \label{f:1diag}}
    \end{center}
\end{figure}

\newpage

\section{New eRASS DR1 CVs presented in this work}\label{a:cvcat}

   \begin{table*}[h]
    \centering 
    \caption{Description of all columns in the online published table of the new CVs identified in this work} 
    \label{t:cat_col}
    \begin{threeparttable}
    \begin{tabular}{lcl}
    \hline
    \hline
    Column name             & Unit   & Description \\  
    \hline
     IAUNAME     &     & Name of system from the eRASS1 catalog\\
     RA\_x  & degrees   & Right ascension in eRASS1 \\
     DEC\_x  & degrees   & Declination in eRASS1 \\
     GaiaID  &   &   Source ID in \gai DR3 \\
     RA\_epoch2000  & degrees   & Right ascension in \gai DR3 \\
     DEC\_epoch2000  & degrees   & Declination in \gai DR3 \\
     Separation\_xo  &  arcsec & Positional offset between X-ray and optical coordinates\\
     $G$  & mag  &   Mean \gai $G$ band magnitude \\
     $BP\_RP$  & mag  &   Difference between \gai $B$-band and      
\gai $R$-band magnitudes \\
    ML\_cts & & Total number of X-ray source photons between 0.2 keV and 2.3 keV \\
     ML\_FLUX  &  erg s$^{-1}$ cm$^{-2}$   & eROSITA DR1 X-ray flux in the $0.2-2.3$ keV band \\     
     Distance   & pc    &   r$_{\text{geo}}$ distance from  \cite{bailer-jones+21}  \\
     $\log L_{\text{X}}$  & erg s$^{-1}$   & Logarithm of the X-ray luminosity in the energy band $0.2-2.3$ keV\\
     $\log(f_{\text{X}}/f_\text{{opt}}$)   &   & Logarithm of f$_{\text{X}}$ to optical using $G_\text{mean}$\tnote{1}\\  
     $G_{\text{abs}}$ & mag & Absolute $G$-magnitude based on $G$ and r$_{\text{geo}}$\\
     SubType    & & Main type that the CV belongs to \\
     Note & & Brief comment on the target \\
     Period & hours & Main period regarded as the orbital period\\
     Period2 & min & Secondary period associated with the spin of the WD or a superhump\\
	\hline
	\hline
    \end{tabular}
    \begin{tablenotes}  
    \item[1] $\log (F_{\text{X}}/F_\text{{opt}}) = \log F_\text{X} + (G/2.5) + 4.86$
    \end{tablenotes}
    \end{threeparttable}
    \end{table*}

\begin{longtable}{|r|l|r|c|l|c|r|}
\caption{Summary of \ero DR1 CV identifications presented in this work. Given are the IAU name from the \ero DR1 catalog, the GaiaID of the optical counterpart, the CV subtype, a brief comment on the object, the primary (likely orbital) period in hours and a secondary period in minutes} \label{t:cvs} \\

\hline 
  \multicolumn{1}{|c|}{seqno} &
  \multicolumn{1}{c|}{IAUNAME} &
  \multicolumn{1}{c|}{GaiaID} &
  \multicolumn{1}{c|}{SubType} &
  \multicolumn{1}{c|}{Note} &
  \multicolumn{1}{c|}{Period} &
  \multicolumn{1}{c|}{Period2} \\
\hline
\endfirsthead

\multicolumn{7}{c}%
{{\bfseries \tablename\ \thetable{} -- continued from previous page}} \\
  \multicolumn{1}{|c|}{seqno} &
  \multicolumn{1}{c|}{IAUNAME} &
  \multicolumn{1}{c|}{GaiaID} &
  \multicolumn{1}{c|}{SubType} &
  \multicolumn{1}{c|}{Note} &
  \multicolumn{1}{c|}{Period} &
  \multicolumn{1}{c|}{Period2} \\
\hline
\endhead

\hline \multicolumn{7}{|r|}{{Continued on next page}} \\ \hline
\endfoot

\hline \hline
\endlastfoot

  1 & 1eRASS J000714.3-533904 & 4923981272433664896 & DN &  &   &  \\
  2 & 1eRASS J000743.4-695943 & 4702994514879977472 & MCV &  & 1.54 &  \\
  3 & 1eRASS J000854.6-870658 & 4613183420122461440 & DN & cand &   &  \\
  4 & 1eRASS J004707.2-484518 & 4926893947455908992 & MCV &  & 1.32 &  \\
  5 & 1eRASS J012009.3-772651 & 4635252400902063872 & MCV &  &   &  \\
  6 & 1eRASS J013330.6-810751 & 4630500071128655360 & Polar & eclipse & 1.54 &  \\
  7 & 1eRASS J014904.7-744845 & 4638176586435463296 & DN &  &   &  \\
  8 & 1eRASS J015011.9-452117 & 4954825592953531392 & DN &  &   &  \\
  9 & 1eRASS J015151.5-473828 & 4942033050042466432 & DN & WZ cand &   &  \\
  10 & 1eRASS J015637.6-835833 & 4617143036371460864 & MCV & cand (NL possible) &   &  \\
  11 & 1eRASS J021344.5-511653 & 4936732888600422400 & DN & cand &   &  \\
  12 & 1eRASS J031621.4-275908 & 5061002548928768128 & MCV & soft X &   &  \\
  13 & 1eRASS J035646.6-444820 & 4835425609500365184 & DN & cand & 2.02 &  \\
  14 & 1eRASS J041429.1-631947 & 4676110149911252992 & DN & cand &   &  \\
  15 & 1eRASS J042405.7-344116 & 4869489888476638080 & DN & cand &   &  \\
  16 & 1eRASS J042633.5-250226 & 4896134903510805888 & DN & cand & 6.62 &  \\
  17 & 1eRASS J044938.1-775603 & 4624507698397085696 & MCV & soft X & 2.35 &  \\
  18 & 1eRASS J045230.9-441103 & 4811451372635865984 & DN &  &   &  \\
  19 & 1eRASS J050325.1-241653 & 2960575279879332480 & DN &  & 2.04 &  \\
  20 & 1eRASS J050330.4+200848 & 3408745152893388800 & DN &  & 1.60 &  \\
  21 & 1eRASS J050740.8-091335 & 3182850042290304768 & Polar & low, Bphot=6-7MG &   &  \\
  22 & 1eRASS J050906.9-032507 & 3214449559796032256 & DN & UG & 8.48 &  \\
  23 & 1eRASS J051939.8+160045 & 3394048084043849728 & MCV & cand &   &  \\
  24 & 1eRASS J052847.6-150300 & 2984105034872059520 & MCV & cand &   &  \\
  25 & 1eRASS J053103.6-010217 & 3220279033010830720 & DN & P2 (TESS) superhump  & 4.52 & 265.20\\
  26 & 1eRASS J054726.8+154051 & 3348188519201136896 & DN & WZ cand &   &  \\
  27 & 1eRASS J055955.0+115559 & 3342232911392616576 & DN & UGSU & 1.49 &  \\
  28 & 1eRASS J060536.3-255946 & 2910450816806553728 & MCV & cand, soft X &   &  \\
  29 & 1eRASS J060637.9-411731 & 5572540506864697856 & DN & UG &   &  \\
  30 & 1eRASS J060652.8+030647 & 3315701047272644352 & DN & cand &   &  \\
  31 & 1eRASS J061056.0+311153 & 3438072155960280704 & NL &  & 7.50 &  \\
  32 & 1eRASS J061712.3+102115 & 3329254314791421056 & Polar & cand & 2.00 &  \\
  33 & 1eRASS J062022.2-052655 & 3008353428092638208 & DN & UGZ/IW &   &  \\
  34 & 1eRASS J064442.1+344710 & 938892254772887040 & DN & UGSU & 1.81 &  \\
  35 & 1eRASS J064501.8-370751 & 5577567504089935488 & DN &  &   &  \\
  36 & 1eRASS J065040.9-072921 & 3099159615233832448 & NL &  &   &  \\
  37 & 1eRASS J065903.9-110033 & 3049105830146397824 & MCV & cand & 5.42 &  \\
  38 & 1eRASS J070038.9-311307 & 5607187247832006656 & DN & cand &   &  \\
  39 & 1eRASS J070801.2-421144 & 5561171556635287680 & MCV & cand &   &  \\
  40 & 1eRASS J071500.5+082309 & 3155137435827580160 & NL & eclipse & 5.07 &  \\
  41 & 1eRASS J071912.0-673423 & 5269056086212803200 & DN & cand &   &  \\
  42 & 1eRASS J071912.6-221840 & 5618220335923507840 & NL &  &   &  \\
  43 & 1eRASS J072515.8-220556 & 5619596095545382784 & NL & cand &   &  \\
  44 & 1eRASS J072615.2-102334 & 3047285653067939584 & DN & UGSS & 10.04 &  \\
  45 & 1eRASS J072920.4+095206 & 3161546596480983040 & CV & MCV or DN &   &  \\
  46 & 1eRASS J074227.4-243915 & 5614672074471772288 & NL & cand, eclipse & 4.19 &  \\
  47 & 1eRASS J074507.7-291032 & 5599061066635017472 & CV & MCV or DN &   &  \\
  48 & 1eRASS J075023.4-481033 & 5518266753197554560 & IP & cand, soft X & 3.20 &  \\
  49 & 1eRASS J075512.7-171456 & 5718184943820583040 & IP &  &   & 33.50\\
  50 & 1eRASS J075522.5+040715 & 3089187354004319104 & DN & cand &   &  \\
  51 & 1eRASS J075753.2-403121 & 5534353609048317184 & MCV & cand &   &  \\
  52 & 1eRASS J080040.0-431105 & 5533020382479508736 & IP &  & 5.39 & 48.00\\
  53 & 1eRASS J080139.0-644402 & 5275656385791353600 & MCV & soft X & 3.60 &  \\
  54 & 1eRASS J081750.3-232201 & 5699480807924198400 & MCV & soft X &   &  \\
  55 & 1eRASS J081916.3-250704 & 5696028032178553728 & MCV &  &   &  \\
  56 & 1eRASS J083228.7-321016 & 5640108966727233280 & DN & cand &   &  \\
  57 & 1eRASS J083420.4-362503 & 5542290124488000256 & NL & cand &   &  \\
  58 & 1eRASS J083533.8-183156 & 5708606204673825408 & DN &  &   &  \\
  59 & 1eRASS J084124.6-263108 & 5646345495463189504 & CV &  &   &  \\
  60 & 1eRASS J084741.0-503019 & 5327813785743801728 & DN &  &   &  \\
  61 & 1eRASS J084937.5-554419 & 5316706347473511936 & MCV & cand &   &  \\
  62 & 1eRASS J090713.0-385902 & 5621267043979633792 & DN & cand &   &  \\
  63 & 1eRASS J090802.4-382811 & 5621307691554972416 & MCV & cand & 3.60 &  \\
  64 & 1eRASS J091300.6-623319 & 5297282649941671168 & DN & cand &   &  \\
  65 & 1eRASS J091312.5-333522 & 5624900689391404160 & MCV & cand &   &  \\
  66 & 1eRASS J092133.6-520350 & 5313179212955281920 & DN & cand &   &  \\
  67 & 1eRASS J092408.8-361540 & 5431265322966878592 & DN &  &   &  \\
  68 & 1eRASS J093244.5-483649 & 5410062203306722048 & DN & cand &   &  \\
  69 & 1eRASS J095507.6-504955 & 5406023525286279680 & DN & cand &   &  \\
  70 & 1eRASS J100139.6-661251 & 5245817770527876992 & DN & cand &   &  \\
  71 & 1eRASS J100328.5-382018 & 5421477436096250368 & DN &  &   &  \\
  72 & 1eRASS J100457.4-541224 & 5356253409930486656 & DN & cand &   &  \\
  73 & 1eRASS J101942.7-263136 & 5471933711157502592 & DN & cand &   &  \\
  74 & 1eRASS J102505.0-730802 & 5229851766470471936 & CV &  &   &  \\
  75 & 1eRASS J103302.8-670141 & 5233150228333079424 & NL & cand &   &  \\
  76 & 1eRASS J103435.4-554041 & 5352208512798624384 & MCV & cand &   &  \\
  77 & 1eRASS J103510.8-560341 & 5352100107825599488 & DN &  &   &  \\
  78 & 1eRASS J103903.9-301246 & 5455143137808991360 & Polar & cand &   &  \\
  79 & 1eRASS J104612.9-511815 & 5360633963010856448 & Polar &  & 1.54 &  \\
  80 & 1eRASS J105334.2-582815 & 5338863293521526016 & CV &  &   &  \\
  81 & 1eRASS J105814.1-561817 & 5340691162878960384 & NL & cand &   &  \\
  82 & 1eRASS J111300.1-174738 & 3558247161967458944 & Polar &  & 3.44 &  \\
  83 & 1eRASS J113928.1-522319 & 5345480326267089792 & IP & cand, soft X &   &  \\
  84 & 1eRASS J114226.5-503132 & 5370060664615498624 & DN & cand &   &  \\
  85 & 1eRASS J120843.8-441209 & 6144540123884994304 & DN & cand &   &  \\
  86 & 1eRASS J122024.6-585041 & 6059398753790788736 & IP & cand &   &  \\
  87 & 1eRASS J123111.9-423518 & 6145666951501477376 & Polar & eclipse & 5.12 &  \\
  88 & 1eRASS J123538.9-532730 & 6077670064363311744 & DN &  &   &  \\
  89 & 1eRASS J123632.5-664601 & 5859908625444588160 & DN & cand &   &  \\
  90 & 1eRASS J123813.9-033935 & 3681313024562519552 & WZ & cand &   &  \\
  91 & 1eRASS J124129.2-605822 & 6056338782906799104 & Polar & eclipse, cycl & 2.22 &  \\
  92 & 1eRASS J125303.4-182755 & 3510296356072166528 & DN &  &   &  \\
  93 & 1eRASS J130120.6-413602 & 6139397368696634240 & DN & UG & 1.55 & 96.20\\
  94 & 1eRASS J130354.3-644710 & 5862002885860903552 & DN &  &   &  \\
  95 & 1eRASS J130814.2-713703 & 5843476492828935040 & Polar & cand, soft X &   &  \\
  96 & 1eRASS J131056.2-402043 & 6140852842917336320 & DN & cand &   &  \\
  97 & 1eRASS J131836.0-710206 & 5843604139272638464 & DN &  &   &  \\
  98 & 1eRASS J132129.9-314108 & 6181357996586920576 & CV & nonmag &   &  \\
  99 & 1eRASS J134035.3-430516 & 6111630130255262336 & IP & cand &   &  \\
  100 & 1eRASS J134212.3-520908 & 6069196884164673152 & DN & cand &   &  \\
  101 & 1eRASS J140146.6-501321 & 6090855029847914368 & IP & cand & 2.17 &  \\
  102 & 1eRASS J141157.3-575942 & 5868116131796083328 & MCV & cand &   &  \\
  103 & 1eRASS J141326.4-694718 & 5846537430115882752 & WZ & cand &   &  \\
  104 & 1eRASS J143410.6-500141 & 5899274062245189888 & MCV &  &   &  \\
  105 & 1eRASS J144156.8-472618 & 5905821035501316224 & NL & cand &   &  \\
  106 & 1eRASS J144858.2-480518 & 5905160847496560384 & Polar & cand, low, Bph=19-20MG &   &  \\
  107 & 1eRASS J145600.4-531821 & 5900004962587152128 & DN & cand &   &  \\
  108 & 1eRASS J145923.8-561402 & 5881264852081787392 & DN & cand &   &  \\
  109 & 1eRASS J150619.5-650333 & 5872888635064556288 & DN & cand &   &  \\
  110 & 1eRASS J150721.5-391052 & 6197581939197319168 & MCV & cand &   &  \\
  111 & 1eRASS J151437.0-531943 & 5888006125342022528 & NL &  &   &  \\
  112 & 1eRASS J151449.8-220059 & 6252460115022163328 & IP & cand &   & 55.00\\
  113 & 1eRASS J151504.8-523611 & 5888318734573513472 & IP & soft X & 5.35 & 26.60\\
  114 & 1eRASS J152211.8-651507 & 5824964015593788800 & NL & cand &   &  \\
  115 & 1eRASS J152620.9-465110 & 5999007326043964928 & DN & cand &   &  \\
  116 & 1eRASS J152721.5-634742 & 5826826274721144064 & MCV & cand &   &  \\
  117 & 1eRASS J152737.6-205306 & 6252119330839755264 & Polar & SO+24 & 1.87 &  \\
  118 & 1eRASS J153936.4-431857 & 6001239231564325376 & IP & cand or NL &   &  \\
  119 & 1eRASS J154026.0-233433 & 6238589363518606464 & MCV &  &   &  \\
  120 & 1eRASS J154103.2-421035 & 6001820662738687104 & MCV & cand, soft X &   &  \\
  121 & 1eRASS J154608.6-324359 & 6015493777024336256 & IP &  &   & 32.40\\
  122 & 1eRASS J154713.8-485316 & 5985902899776646784 & NL &  &   &  \\
  123 & 1eRASS J155226.7-605844 & 5832685885740170368 & DN & cand (AT2023sev) &   &  \\
  124 & 1eRASS J155627.5-305153 & 6039779205745065984 & DN &  &   &  \\
  125 & 1eRASS J155901.2-443809 & 5994311999047362688 & NL &  & 3.69 &  \\
  126 & 1eRASS J160209.4-423052 & 5994940674888088832 & Polar & cand & 1.99 &  \\
  127 & 1eRASS J161313.0-710118 & 5807187661316522240 & Polar & cand, soft X &   &  \\
  128 & 1eRASS J161340.4-292858 & 6038652760384797824 & MCV & cand &   &  \\
  129 & 1eRASS J161355.3-304640 & 6038254840256913792 & MCV & IP cand &   &  \\
  130 & 1eRASS J162732.2-373334 & 6018793136543095424 & Polar & cand &   &  \\
  131 & 1eRASS J163630.6-504655 & 5940264542959110656 & NL & VY Scl cand & 1.96 &  \\
  132 & 1eRASS J164623.7-194916 & 4130489496180445184 & IP & cand & 5.78 &  \\
  133 & 1eRASS J164641.8-421318 & 5967960514965805696 & NL & cand &   &  \\
  134 & 1eRASS J170538.1-292322 & 6029802344643324800 & Polar & cand &   &  \\
  135 & 1eRASS J171617.5-362418 & 5973940144536293376 & DN & UG & 21.78 &  \\
  136 & 1eRASS J171755.6-580619 & 5916322952093141120 & DN & WZ cand &   &  \\
  137 & 1eRASS J172757.7-324306 & 4054744522392393216 & NL & cand &   &  \\
  138 & 1eRASS J173252.9-440727 & 5958345771178035200 & IP &  &   & 33.40\\
  139 & 1eRASS J173706.6-525042 & 5922000795798229888 & NL & cand &   &  \\
  140 & 1eRASS J173836.5-340837 & 4053478606615162752 & MCV & cand &   &  \\
  141 & 1eRASS J174508.8-505151 & 5946454415417964032 & MCV & cand, Zeeman &   &  \\
  142 & 1eRASS J175312.5-451037 & 5954644948117588352 & DN &  &   &  \\
  143 & 1eRASS J175455.8-443323 & 5954589663316137472 & Polar & Bcycl=47MG &   &  \\
  144 & 1eRASS J180106.4-454404 & 6720591273752642176 & NL & VY Scl cand &   &  \\
  145 & 1eRASS J180549.1-511557 & 6702812656064157952 & MCV & cand & 3.00 &  \\
  146 & 1eRASS J183256.2-393751 & 6723806864218464256 & DN & cand &   &  \\
  147 & 1eRASS J184412.7-424652 & 6710261774316684672 & Polar & cand & 2.06 &  \\
  148 & 1eRASS J190459.8-652424 & 6435381393471165312 & Polar & cand, soft X &   &  \\
  149 & 1eRASS J190531.9-534943 & 6644349801742495488 & DN &  & 1.55 &  \\
  150 & 1eRASS J191213.8-441044 & 6712706405280342784 & WD pulsar & IP+23, AS+23 &   &  \\
  151 & 1eRASS J193934.9-642156 & 6440610262752704896 & Polar & soft X & 2.31 &  \\
  152 & 1eRASS J195506.0-495744 & 6670044490834142080 & DN & cand &   &  \\
  153 & 1eRASS J211136.0-434137 & 6579892337013464320 & DN &  &   &  \\
  154 & 1eRASS J214354.9-572038 & 6458951727315217152 & DN & cand, eclipse &   &  \\
  155 & 1eRASS J230721.4-400033 & 6543375597350678656 & Polar & eclipse, soft X & 2.07 &  \\
  156 & 1eRASS J234504.4-430343 & 6532165251671628416 & DN & cand &   &  \\
\hline
\end{longtable}
Notes: SO+24 -- \cite{ok+24}; IP+23 -- \cite{pelisoli+23}; AS+23 -- \citep{schwope+23}

\section{Spectroscopic identification of further \ero X-ray sources}\label{a:other}
Table \ref{t:oxid} lists spectroscopic identifications of \ero X-ray sources that were targeted as CV candidates but were identified mostly as AGN. The table lists their IAU names from the DR1 catalog, the \gai source ID, the separation between the X-ray and the optical coordinates, and the type of X-ray emitters. The likely stellar counterparts are discussed individually below. In Fig.~\ref{f:oxid} diagnostic diagrams are shown that demonstrate the similarity between the CVs and the contaminating sources. For the AGN this is simply due to their non-recognized extragalactic nature from \gai.

1eRASS J100546.0-365326: A spectrum was obtained with NTT/EFOSC2 of \gai DR3 5422122089209249792, $G=19.0, BP-RP=0.52, r_{\rm geo} = 1010$\,pc, which was classified as an M-star without obvious signs of activity (no H$\alpha$ or CaHK emission lines). For such a star the inferred X-ray to optical flux ratio and X-ray luminosity are too large, $f_{\rm X}/f_{\rm opt} = -0.24$ and $\log L_{\rm X} = 31.4$\,\lx. A second object, \gai DR3 5422122089211774208, 4.5 arcsec offset from the DR1 X-ray position, was later (while reducing the spectroscopy run) identified also as a possible counterpart, but no spectrum could be taken. If this would be the correct counterpart, its distance, flux ratio and luminosity would be compatible with a CV nature. \gai DR3 5422122089209249792 is also listed in the counterpart catalog of \cite{salvato+25} but is excluded here. The observed star is shown with red symbols, the other possible counterpart with cyan symbols in Fig.~\ref{f:oxid}.

\begin{table*}[h]
  \centering
  \caption{Further \ero identifications\label{t:oxid}}
\begin{tabular}{|l|r|r|l|l|}
\hline
  \multicolumn{1}{|c|}{IAUNAME} &
  \multicolumn{1}{c|}{GaiaID} &
  \multicolumn{1}{c|}{Separation ["]} &
  \multicolumn{1}{c|}{CLASS} &
  \multicolumn{1}{c|}{redshift} \\
\hline
  1eRASS J230912.5-424633 & 6541754882850907264 & 3.5 & AGN & 0.631\\
  1eRASS J015436.6-505242 & 4937478567939069056 & 5.0 & AGN & 0.190\\
  1eRASS J042617.3-592328 & 4678800968397318016 & 1.8 & AGN & 0.103\\
  1eRASS J092940.9-341113 & 5437912008196121472 & 2.9 & AGN & 0.112\\
  1eRASS J111111.2-463022 & 5374859826706510592 & 2.9 & AGN & 0.086\\
  1eRASS J100546.0-365326 & 5422122089209249792 & 1.7 & STAR M& \\
  1eRASS J100546.0-365326 & 5422122089211774208 & 4.6 & CV cand &\\
\hline\end{tabular}
\end{table*}

\begin{figure*}[h]
\resizebox{0.5\hsize}{!}{\includegraphics[clip=]{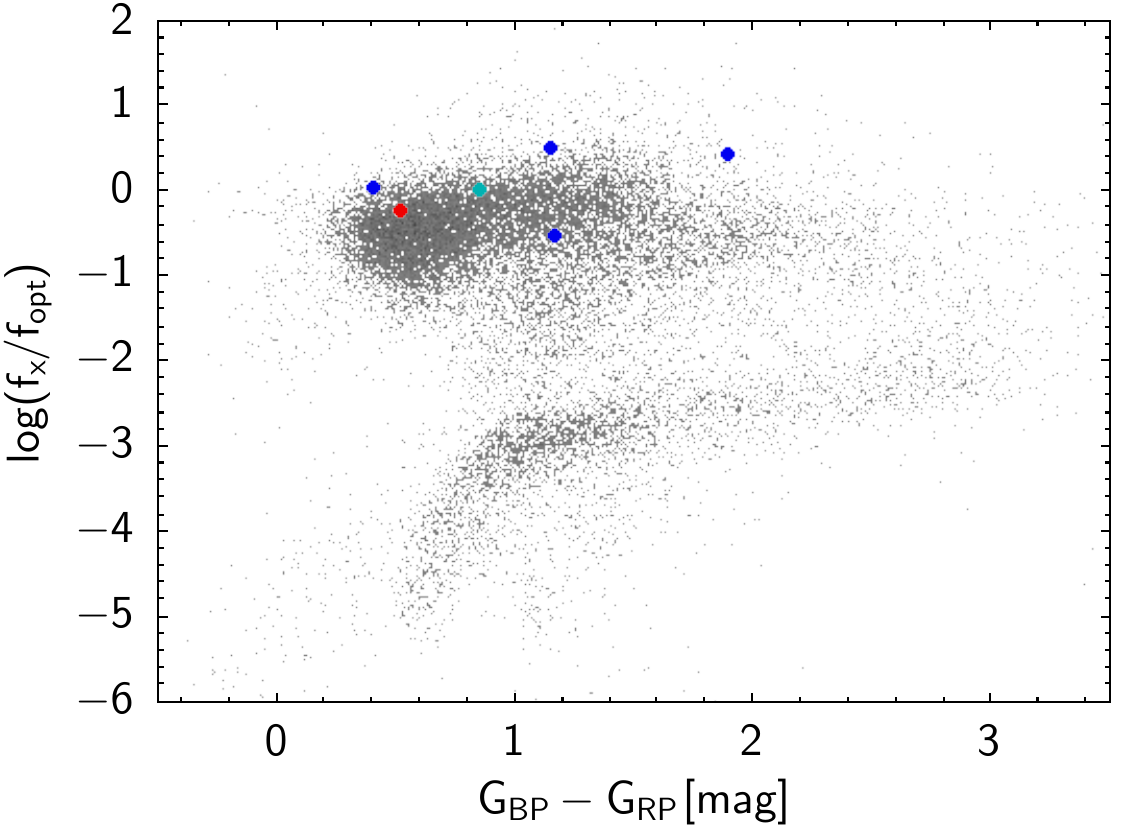}}
\resizebox{0.5\hsize}{!}{\includegraphics[clip=]{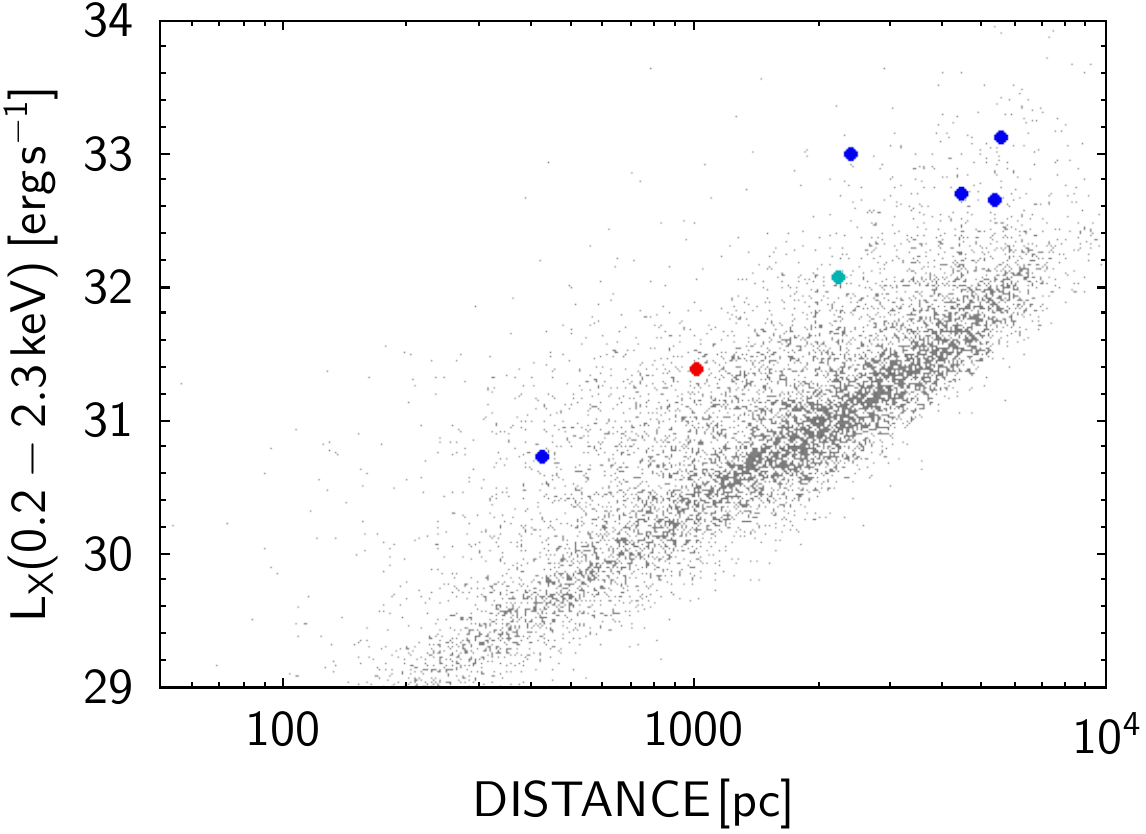}}
\resizebox{0.5\hsize}{!}{\includegraphics[clip=]{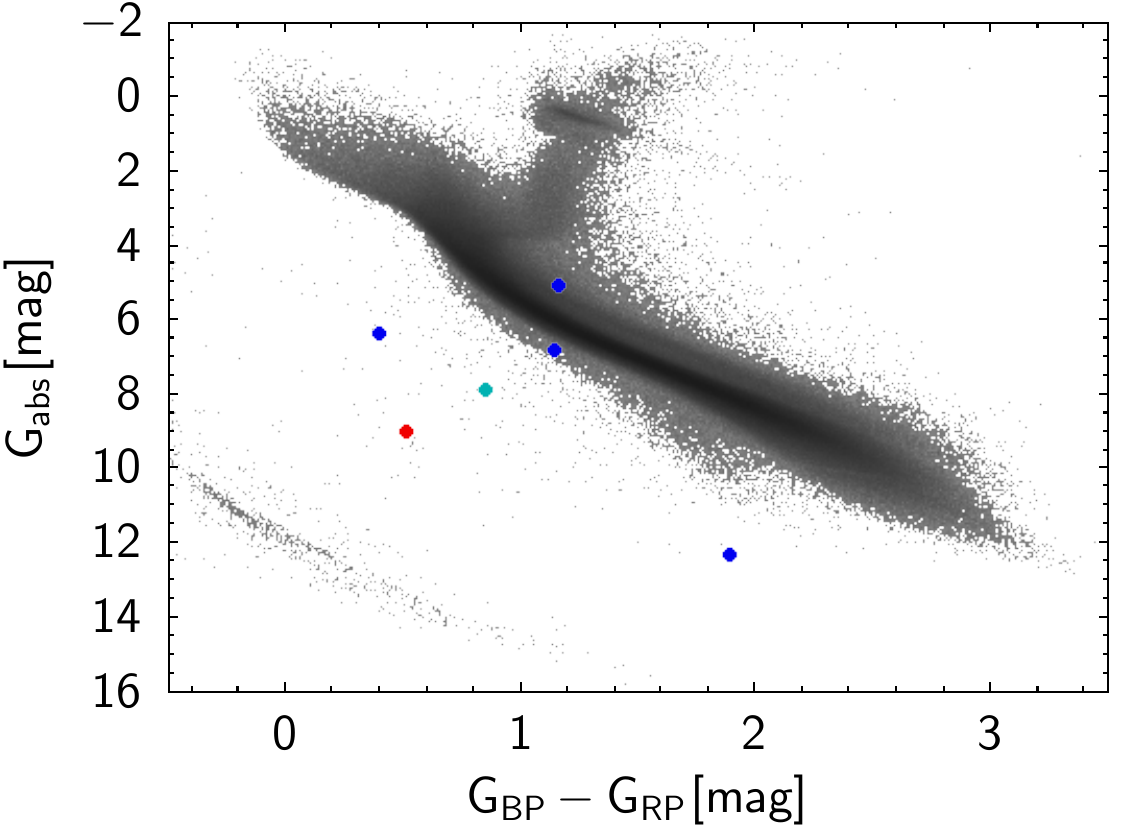}}
\resizebox{0.5\hsize}{!}{\includegraphics[clip=]{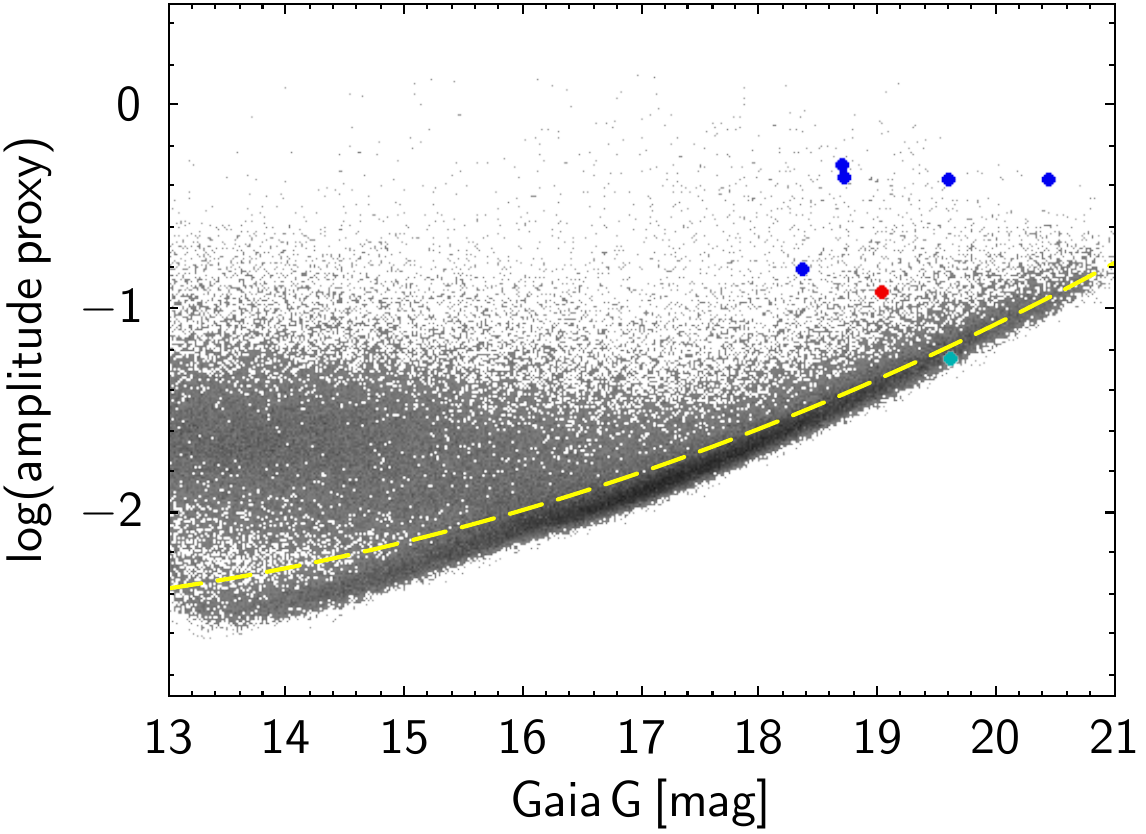}}
\caption{Diagnostic diagrams of confirmed X-ray counterparts (AGN, blue symbols) and the candidate objects of 1eRASS J100546.0-365326 (see discussion above).
\label{f:oxid}
}
\end{figure*}

\end{appendix}
\end{document}